\documentclass[journal]{IEEEtran}

\ifCLASSINFOpdf
\else
   \usepackage[dvips]{graphicx}
\fi
\usepackage{url}

\usepackage{amsmath,amssymb,amsthm,graphicx,amsfonts,amsbsy,latexsym,booktabs}

\graphicspath{{figs/}}

\usepackage[noadjust]{cite}

\usepackage{graphicx}
\usepackage{epstopdf}
\usepackage{epsf}
\usepackage{caption}
\usepackage{epsfig}

\newcommand{\Real}{\operatorname{Re}}

\DeclareMathOperator*{\argmin}{argmin}
\DeclareMathOperator*{\minimize}{minimize}

\newtheorem{citedthm}{Theorem}

\newtheorem{citedprop}[citedthm]{Proposition}

\newcommand{\R}{\mathbb R}

\newcommand{\ra}[1]{\renewcommand{\arraystretch}{#1}}

\begin{document}

\title{RIP sensing matrices construction for sparsifying dictionaries with application to MRI imaging}

\author{J. Ho, W-L. Hwang, and A. Heinecke, 
\thanks{J. Ho and W.-L. Hwang are with Academia Sinica, Taipei 11529, Taiwan (e-mail: whwang@iis.sinica.edu.tw).}
\thanks{A. Heinecke is with the National University of Singapore, 119077 Singapore.}}

\markboth{Journal of \LaTeX\ Class Files, Vol. 14, No. 8, August 2015}
{Shell \MakeLowercase{\textit{et al.}}: Bare Demo of IEEEtran.cls for IEEE Journals}
\maketitle

\begin{abstract}
Practical applications of compressed sensing often restrict the choice of its two main ingredients. They may (i) prescribe the use of particular redundant dictionaries for certain classes of signals to become sparsely represented, or (ii) dictate specific measurement mechanisms which exploit certain physical principles. On the problem of RIP measurement matrix design in compressed sensing with redundant dictionaries, we give a simple construction to derive sensing matrices whose compositions with a prescribed dictionary have with high probability the RIP in the $k \log(n/k)$ regime. Our construction thus provides recovery guarantees usually only attainable for sensing matrices from random ensembles with sparsifying orthonormal bases. Moreover, we use the dictionary factorization idea that our construction rests on in the application of magnetic resonance imaging, in which also the sensing matrix is prescribed by quantum mechanical principles. 
We propose a recovery algorithm based on transforming the acquired measurements such that the compressed sensing theory for RIP embeddings can be utilized to recover wavelet coefficients of the target image, 
and show its performance on examples from the fastMRI dataset.
\end{abstract}


\IEEEpeerreviewmaketitle

\section{Introduction} \label{intro}

\IEEEPARstart{C}{ompressed}  sensing (CS)  provides a framework under which sparse or compressible signals can be stably reconstructed from far fewer linear measurements than their ambient dimension \cite{candes2006robust, donoho2006compressed}. 
The number of required measurements depends on the signal complexity in terms of sparsity, and properties of the sensing matrix with respect to sparse vectors, such as the restricted isometry property (RIP) \cite{candes2005decoding}. 
A sensing matrix $S$, which embeds high-dimensional signals into a lower-dimensional measurement space, is said to have the RIP of order $k$ if there exists a constant $\delta_k \in [0, 1)$, such that for any $k$-sparse vector $x \in \R^n$,
\begin{align} \label{SRIP}
(1 - \delta_k) \| x \|_2^2 \leq \|Sx \|_2^2 \leq (1 + \delta_k) \|x \|_2^2.
\end{align}
Improving upon~\cite{candes2006stable},
Cand\`{e}s \cite{candes2008restricted} showed that $k$-sparse and compressible vectors can be stably recovered from observations $y = Sx + \eta$ with measurement error $\eta$ bounded by $\epsilon$, given that $S$ has the RIP with  $\delta_{2k} < \sqrt 2 - 1$ (which has been improved to $0.4652$~\cite{Foucart2010RIC}) via the sparsity promoting convex program
\begin{align}\label{CSgeneral}
\minimize_x \|x\|_1 \quad \text{subject to } \quad \|y-Sx\|_2\leq\epsilon.
\end{align}
For certain $\lambda>0$, \eqref{CSgeneral} can be equivalently reformulated to the problem of minimizing over $x$ the unconstrained objective $\tfrac{1}{2}\|y-Sx\|_2^2+\lambda \|x\|_1$ consisting of data-fidelity and regularization term, which can be solved using iterative methods~\cite{Tropp2006JustRelax}.

While it is NP-hard to verify the RIP for a given matrix,
random matrices from subgaussian or Bernoulli ensembles do possess the RIP with high probability - also when  composed with  orthonormal bases \cite{baraniuk2008simple}. 
In practice, however, many classes of signals are sparse only with respect to a redundant dictionary or tight frame that is non-orthonormal (e.g., Gabor, curvelet, wavelet or data-driven learned dictionaries). In this case, the sensing matrix $S$ in \eqref{SRIP} has to be replaced by the composition $SD$, which no longer possesses the RIP. This prevents the RIP as recovery guaranteeing tool in many CS applications.

An important CS applications, in which the sensing mechanisms is dictated by hardware constraints, is magnetic resonance imaging (MRI). In MRI a scanner collects Fourier domain measurements of a target medical image. Accelerating the speed of MRI data acquisition by reducing the number of required measurements remains of great interest to the medical community. 
Inference of the true underlying spatial image can be achieved via several CS strategies when non-uniformly undersampling below the requirements of the Shannon-Nyquist theory within the maximum frequency spectrum, by incorporating the a-priori knowledge about sparsity of medical images in a dictionary transform domain or of its spatial gradients \cite{Lustig2007}. 
While deep learning approaches have been a recent centre of attention to recover data undersampled in this way \cite{Pal2022DLMRIreview}, recovery guarantees and interpretability set the CS approach apart from such machine learning techniques \cite{gu2022revisiting}.
Though state-of-the-art machine learning models for MRI have been validated for clinical interchangeability in 4-fold data aquisition acceleration~\cite{Recht2020Interchangeability}, they rely on empirical studies and present challenges concerning, for instance, reconstruction hallucinations that can be problematic for interpreting radiologists~\cite{fastMRI:kspaceandDICOM2020}, unknown training data biases~\cite{Shimron2021SubtleInverseCrimes}, or robustness to distribution shifts between training and application data (e.g., scanning technology~\cite{Liu2022UndersampledMR}, target anatomy~\cite{Liu2021MRIAnatomyTransference}, or acceleration factors
\cite{Taori2020DistributionShifts}). 
In contrast, the sparsity-driven CS-approach can be fine-tuned to perform close to deep learning methods for MRI acceleration, via un-rolling algorithms which consist of a small fraction of the parameters employed by deep learning approaches~\cite{gu2022revisiting}.

\emph{Contributions:} We first present a method to derive a sensing matrix $S \in \R^{m \times l}$ for a given sparsifying dictionary $D \in \R^{l \times n}$, such that with high probability $SD$ has the RIP for $m$ on the order of $k \log (n/k)$, where $k$ is the sparsity level of a coefficient vector. We use a sufficient condition for the RIP of random matrices that satisfy a concentration of measure inequality \cite{baraniuk2008simple}. Starting from any random matrix $A$ of the dictionary dimension for which a random row-selection $\mathcal EA \in \R^{m \times n}$ satisfies this concentration inequality, a tailored sensing matrix can be derived from any factorization $D = GAH$ in which $G \in \R^{l \times l}$ is invertible and $H \in \R^{n \times n}$ is orthonormal: For the sensing matrix $S := \mathcal EG^{-1}$ the composition $S D$ has then the desired RIP with high probability. 
We show in  Sect.~\ref{mainsec} that the required factorization exists whenever $D$ and $A$ have equal rank, and detail constructions. A particular implication is that one can thus obtain the same RIP-based recovery guarantees for sensing matrix compositions with general over-complete dictionaries as for Gaussian or Bernoulli ensembles with orthonormal bases.

We then apply the dictionary factorization idea in the CS MRI application. In CS MRI, sensing mechanisms exploit quantum mechanical principles and existing hardware constraints restrict sensing matrix design beyond non-uniformly subsampling the measured (complex-valued) Fourier spatial frequency coefficients of the target image $\widetilde x=Dx$, which we suppose is synthesised from (real-valued) sparse coefficients $x$ with respect to a dictionary $D$. The sensing matrix $S=\mathcal{R}F$ is thus modelled as product of a discrete Fourier transform $F$ and a row subsampling $\mathcal{R}$.
The CS recovery of $x$ can then be formulated as the \emph{$\ell_1$-synthesis} problem
\begin{align} \label{MRIsyn}
\minimize_{x} \| x\|_1 \quad\text{ subject to }\quad \| y -  \mathcal{R}F D x \|_2 \leq \epsilon
\end{align}
in which $\mathcal{R}F$ possesses the RIP ~\cite{CandesTao2006NearOptimal,RudelsonVershynin2008,Bourgain2014RIP}, while, in general, $\mathcal{R}F D$ does not, such that the RIP recovery guarantees do not directly apply.
%
Using the dictionary factorization idea we attempt to neutralize the effect of $F$ in order to obtain an RIP synthesis problem. We do so by choosing (real) factors $G$ such that (both real and imaginary parts of) $\mathcal{R}FG$ optimally match $\mathcal{R}$, so that the factorizations $GAH$ that optimally approximate $D$ allow, by virtue of $\mathcal{R}FD\sim\mathcal{R}FGAH\sim\mathcal{R}AH$, to replace the constraint in \eqref{MRIsyn} by $\| y -  \mathcal{R}AHx \|_2 \leq \epsilon$.
Both the real and the imaginary parts of $\mathcal{R}A H$ are incoherent sampling matrices for sparse signals that possess the RIP. 
We detail this application to MRI in 
Sect.~\ref{mrisec}, and report in Sect.~\ref{experiment} numerical experiments comparing our approach to total-variation based CS MRI \cite{Donoho2008CSMRI_Intro}. 


\emph{Related work:}
If the sparsifying dictionary in the general $\ell_1$-synthesis approach is a tight frame, i.e., $x=D^\top \widetilde{x}$, recovery guarantees can be derived by considering the $\ell_1$-analysis approach of minimizing over $\widetilde{x}$ the objective
$\|D^\top\widetilde{x}\|_1$ subject to the constraint $\|y-S\widetilde{x}\|_2\leq \epsilon$.
Recovery guarantees for the $\ell_1$-analysis approach are connected to the notion of D-RIP~\cite{candes2011compressed}, which requires the sensing matrix to satisfy the RIP inequality for all images of $k$-sparse vectors under a tight frame $D$.  
Any RIP-matrix satisfies the D-RIP when multiplied by a random sign matrix~\cite{KramerWard2011}. 
Unless $D$ is orthonormal, the geometric structures, properties and empirical performances of the $\ell_1$-analysis and $\ell_1$-synthesis approaches in general differ~\cite{Elad2007AnalysisVsSynthesis}. 

A weaker condition than the RIP that can facilitate recovery guarantees for $SD$ with general dictionary $D$ is the mutual incoherence between the sensing matrix and the dictionary~\cite{tropp2007signal,donoho2006compressed}. 
However, sparsity ranges for signals to be recovered, as well as necessary number of measurements are more restrictive than for RIP guarantees, and overcomplete dictionaries with highly coherent columns in general lead to large coherence of their product with sensing matrices. 
%
%
Finally, the nullspace property of a sensing matrix gives a necessary and sufficient condition for stable recovery of sparse signals using the convex optimization \cite{cohen2009compressed}.



\section{Sensing matrix construction} \label{mainsec}

Signal complexity, in terms of sparsity, determines the amount of possible undersampling in CS. 
Constructions utilizing randomness can produce RIP matrices for which the number of required measurements $m$ scales linearly with the sparsity level $k$ of the vector to be recovered. Such matrices can be derived from distributions for which the following concentration of measure inequality, resembling \eqref{SRIP}, holds, such as for subgaussian or Bernoulli ensembles.
\begin{citedthm}[\cite{baraniuk2008simple}] \label{Lem:Baraniuk}
Let $0 < \delta_k < 1$ and $A \in \mathbb R^{m \times n}$ be an iid random matrix. If $\mathbb E\|A x \|_2^2 = \| x\|_2^2$ for all $x \in \R^n$, and for any $\epsilon \in (0, 1)$ the concentration inequality $\mathbb{P}(\|Ax\|_2^2 - \|x\|_2^2| \geq \epsilon \|x\|_2^2) \leq 2 e^{- l c(\epsilon)}$ holds for some $c(\epsilon)>0$ and all $x \in \R^n$, then there exists $c_1, c_2 > 0$ (depending only on $\delta_k$) such that, whenever $k \leq \frac{c_1m}{ \log(n/k)}$, the RIP (\ref{SRIP}) holds for $A$ with probability at least $1-2 e^{-c_2 m}$. 
\end{citedthm}

Random matrices satisfying the concentration inequality are universal with respect to orthonormal bases~\cite{baraniuk2008simple}, i.e., the same RIP conclusions hold for their products with unitary matrices. We use this universality to derive sensing matrices for any sparsity inducing dictionary by random row selection of an invertible transform adapted to the dictionary:

\begin{citedthm} \label{Thm:Main}
Let $D \in \mathbb{R}^{l \times n}$ ($l\leq n$) be a dictionary, $A\in \mathbb{E}^{l \times n}$ and $\mathcal{E}\in\mathbb{E}^{m\times l}$ matrices such that $\mathcal{E}A$ satisfies the assumptions of Theorem~\ref{Lem:Baraniuk}. Suppose the dictionary allows a factorization $D=GAH$ for some invertible $G$ and orthonormal $H$.
Then $S:=\mathcal{E}G^{-1} \in \R^{m \times l}$ is a sensing matrix for $D$ such that, with probability as in Theorem~\ref{Lem:Baraniuk}, $SD$ has the RIP whenever $m\gtrsim k \log (n/k)$. 
\end{citedthm}

In this construction $SD$ equals $\mathcal{E}AH$, which implies the claim due to the above mentioned universality.  
The existence of a factorization $D=GAH$ as assumed in Theorem~\ref{Thm:Main} requires $A$ and $D$ to have equal rank. We next show that the latter also suffices. Given a full-rank sparsifying dictionary, the sensing matrix construction of Theorem~\ref{Thm:Main} can therefore (with probability one) be carried out, starting from a Gaussian matrix $A$ of the same dimensions as the dictionary, and a random row selection $\mathcal{E}$.

\begin{citedprop}\label{prop1}
Let $A,D \in \mathbb{R}^{l\times n}$ with $l \leq n$. Then the following statements are equivalent: 
\begin{itemize}
\item[(i)] There exists an invertible
$G\in \mathbb{R}^{l\times l}$ and  orthonormal $H\in \mathbb{R}^{n\times n}$ such that
$GD = AH$.
\item[(ii)]  There exists an invertible
$G\in \mathbb{R}^{l\times l}$ with
$GDD^\top G^\top = AA^\top $.
\item[(iii)]
$A$ and $D$ have equal rank.
\end{itemize}
\end{citedprop}
\begin{IEEEproof}
We show that (iii) implies (ii), and in turn (i). Assuming (iii), $AA^{\top}$ and $DD^{\top}$ have equal rank. 
Their respective spectral decompositions
$Q_A  \Sigma_A Q_A^{\top}$ and
$Q_D  \Sigma_D Q_D^{\top}$
can thus be assumed to have zeros in the same positions of the diagonal matrices $\Sigma_A$, $\Sigma_D$. Replacing those zero diagonal entries by ones to result in $\Sigma'_A$, $\Sigma'_D$,
the matrix
$G:=Q_A  (\Sigma'_A\Sigma_D^{'-1})^{1/2}Q_D^{\top}$
is invertible and
\begin{align*}
GDD^{\top} G^{\top}
&=G Q_D  \Sigma_D Q_D^{\top} G^\top \\
&= Q_A  (\Sigma'_A\Sigma_D^{'-1})^{1/2} \Sigma_D (\Sigma'_A\Sigma_D^{'-1})^{1/2} Q_A^{\top} = AA^{\top}. 
\end{align*}
Defining $H:=A^+GD+N_AN_D^\top$,
with $A^+$ denoting the pseudo-inverse of $A$, and $N_A$, resp.\ $N_D$, having pairwise orthonormal columns spanning the nullspaces of $A$, resp.\ $D$, 
implies
\[
AH = AA^+GD+AN_AN_D^\top =  AA^+GD = GD
\]
since $AA^+$ is the identity on the range of $A$, which contains the range of $G$. Moreover,
\begin{align*}
HH^\top &= A^+GDD^\top G^\top(A^+)^\top + N_AN_A^\top \\
&= 
A^+  A A^\top (A^+)^\top +  N_AN_A^\top = 
A^+ A + N_A N_A^\top,
\end{align*}
which is the identity.
Finally, note that (i) implies (iii).
\end{IEEEproof}

\medskip
An alternative factorization can be derived from orthonormal bases for the ranges and nullspaces of prescribed equal rank $A,D \in \mathbb{R}^{l\times n}$ as follows. Let $[U_A \,\, N_A],[U_D \,\, N_D]$ be orthonormal matrices, where the columns of $U_A,U_D$ span the ranges, and the columns of $N_A,N_D$ the nullspaces, of $A,D$. 
Extend $DU_D, AU_A$ to invertible square matrices $\widetilde{DU_D},\widetilde{AU_A}$, appending columns. Then
\begin{align}\label{rangeConstructions}
G :=\widetilde{DU_D}\widetilde{AU_A}^{-1} 
\quad\text{ and }\quad
H := U_AU_D^{\top} + N_AN_D^\top
\end{align}
are invertible, respectively orthonormal, and $GAH-D=(GAU_A-DU_D)U_D^{\top}$ is zero since $GAU_A=DU_D$.

\medskip
We next detail a factorization for the practically important case of a prescribed tight frame dictionary $D$. We may then assume $DD^\top$ to be the identity.

\begin{citedthm} \label{tightframe}
Suppose $A,D \in \mathbb{R}^{l\times n}$ have full rank $l\leq n$, and $D$ is a tight frame. Then $D=GAH$ with invertible, resp.\ orthonormal,
\begin{align*}
G := O(AA^\top)^{-1/2} 
\quad\text{ and }\quad
H := A^\top G^\top D +  N_A N_D^\top,
\end{align*}
where $O\in\R^{l\times l}$ is any orthonormal matrix, and the columns of $N_A , N_D\in\R^{n\times{n-l}}$ are any orthonormal bases of the nullspaces of $A$, resp.\ $D$. In particular, if $A$ is also a tight frame, then $G$ is orthonormal.  
\end{citedthm}
\begin{IEEEproof}
Since $GAA^{\top}G^\top$ is the identity and $A N_A$ the zero operator, a direct calculation shows $GAH = D$. While $G$ is by definition invertible, 
$DD^\top$ being the identity 
implies that
$HH^{\top} 
=  A^{\top} (AA^{\top})^{-1} A +N_AN_A^{\top} 
$
is the identity.
\end{IEEEproof}


\section{Application to accelerated MRI imaging} \label{mrisec}

The goal of fast MRI is the recovery of an image $Z\in \mathbb{R}^{n_1 \times n_2}$ from undersampled Fourier ($k$-space) measurements.
The discrete Fourier transform of $Z$ is the complex matrix
\begin{align*}
\widehat{Z} := F_1 Z F_2 \in \mathbb{C}^{n_1 \times n_2},
\end{align*}  
where $F_1 \in \mathbb{C}^{n_1 \times n_1}$, $F_2 \in \mathbb{C}^{n_2 \times n_2}$ are (symmetric and unitary) discrete Fourier transform matrices. Data acquired via accelerated MRI can be modelled as 
\begin{align} \label{undersampling}
Y := \mathcal{R} \widehat{Z} \in \mathbb{C}^{n_1 \times n_2},
\end{align}
where $\mathcal{R}$ is a $\{0,1\}$-entry $n_1\times n_1$ diagonal matrix that selects rows of $Y$ corresponding to some undersampling scheme~\cite{zbontar2018fastmri}.
The image can be assumed to have a sparse representation 
\begin{align}\label{decompose_im}
Z = D_1 X D_2^\top
\end{align} 
in the transform domain of wavelet dictionaries $D_i \in \mathbb{R}^{n_i \times N_i}$ ($N_i\geq  n_i$, $i=1,2$) that implement a separable bivariate wavelet transform. The coefficient matrix $X\in \mathbb{R}^{N_1 \times N_2}$ then consists of low-pass channel approximation coefficients, and sparse high-pass channel coefficients that capture edges and singular points in the target image.
In this formulation, the CS-MRI problem is then the inverse problem of approximating sparse coefficients $X$ that solve
\begin{align}\label{measurements}
Y=\mathcal{R}F_1D_1XD_2^\top F_2,
\end{align}
i.e., to reconstruct $Z$ (via \eqref{decompose_im}) from the measurements $Y$
which are in general corrupted by noise.

Being tied to the measurements \eqref{measurements} by physical principles, we first transform the acquired data. Starting from a sensing factorization $D_2=G_2AH_2$ according to Proposition~\ref{prop1}, we consider the transformed observation
\[
\widetilde{Y} := Y F_2^*(G_2^\top)^{-1}
=\mathcal{R}F_1D_1X(A_2H_2)^\top.
\]
We next construct for $D_1$ two (real) factorizations that are adapted to the real and imaginary parts of $F_1$. First, we let
\begin{align}\label{opti_G}
G_R := \argmin_{G\in\R^{n_1\times n_1} \text{ invertible}}  \left\| \mathcal{R}(\Real(F_1)G -I)\right\|_F^2 
\end{align}
and, second, we let
\begin{align}\label{opti_H}
H_R := \argmin_{H\in\R^{n_2\times n_2} \text{ unitary}}  \left\| D_1-G_RAH\right\|_F^2.
\end{align}
Analogously we define $G_I$ and $H_I$ via solving \eqref{opti_G} for the imaginary part of $F_1$, before solving \eqref{opti_H} for $G_I$. We then let $H:=H_R+iH_I$.
For details on numerical solutions of \eqref{opti_G} and \eqref{opti_H} via the augmented Lagrangian approach we refer to the Appendix.

Our motivation for desiring in \eqref{opti_G}  small differences between $\mathcal{R}$ and the real (resp.\ imaginary) parts of $\mathcal{R}F_1G_R$ (resp. $\mathcal{R}F_1G_I$) is that then \eqref{opti_H} implies that the difference 
%
\begin{align*}
\Phi(X) :=\tfrac{1}{2}\left\|\widetilde{Y}
-
\mathcal{R}  A H
X (A H_2)^\top\right\|^2_F
\end{align*}
is small, such that we can attempt to approximate sparse coefficients $X$ by solving the $\ell_1$-regularized problem
\begin{align}\label{SparseApprox}
\minimize_{X\in\R^{N_1\times N_2}}
\Phi(X)
+ \gamma\| X^{H} \|_1,
\end{align}
where $X^H$ denotes the highpass coefficients of $X$, and $\gamma>0$.

\section{Experimental results} \label{experiment}

\subsection{Sensing matrix of sparsifying dictionaries}

Experiments were conducted to assess the sensing matrix derived based on the dictionary factorizations proposed in Sect.~\ref{mainsec} for CDF 9/7 wavelets and dictionaries derived via K-SVD, for which synthetically generated univariate sparse signals are then shown to be recoverable with the guarantees of Theorem~\ref{Thm:Main}.

\begin{figure}[h!]
{\centerline{\hspace{-0.9cm}
\epsfig{figure=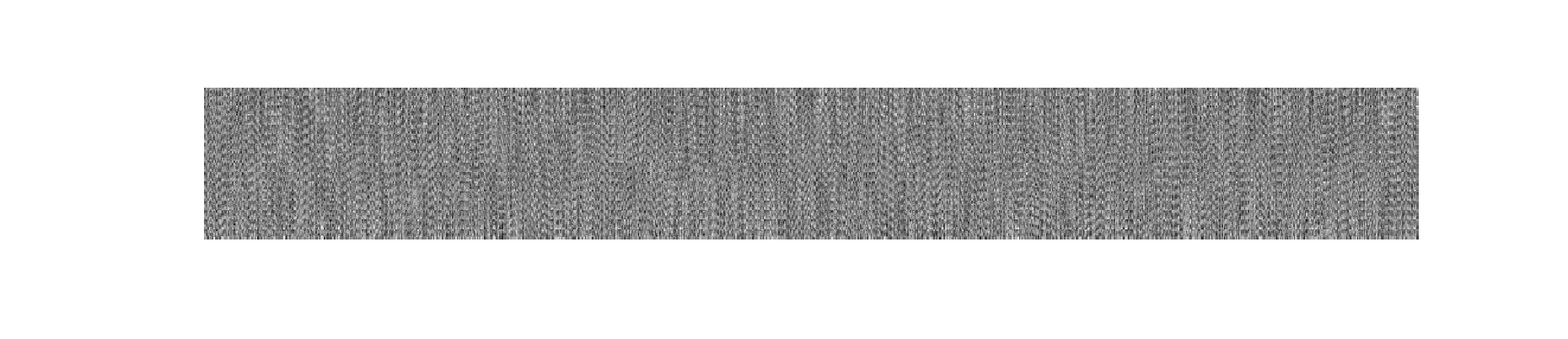,width=9.5cm}}}
\centerline{\hspace{0.8cm}(a) $D_{KSVD}$}
{\centerline{\hspace{-0.9cm}
\epsfig{figure=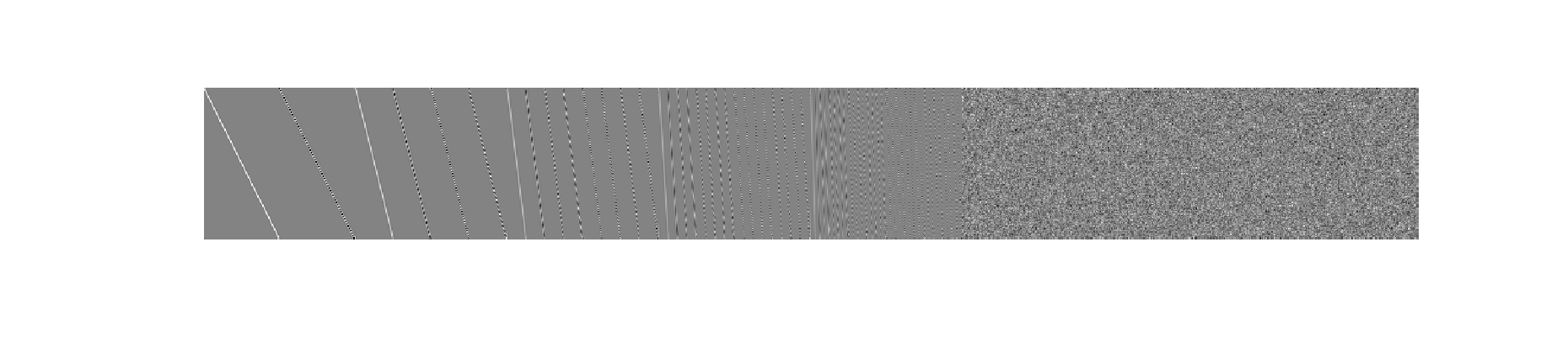,width=9.5cm}}}
\centerline{\hspace{0.8cm}(b) $D_{wavelet}$ }
\caption{\label{fig_D}Visualization of sparsifying dictionaries (of size $128 \times 1024$) with integer values ranging from $0$ to $255$.}
\end{figure}

\begin{figure}[h!]
{\centerline{\hspace{0.0cm}\epsfig{figure=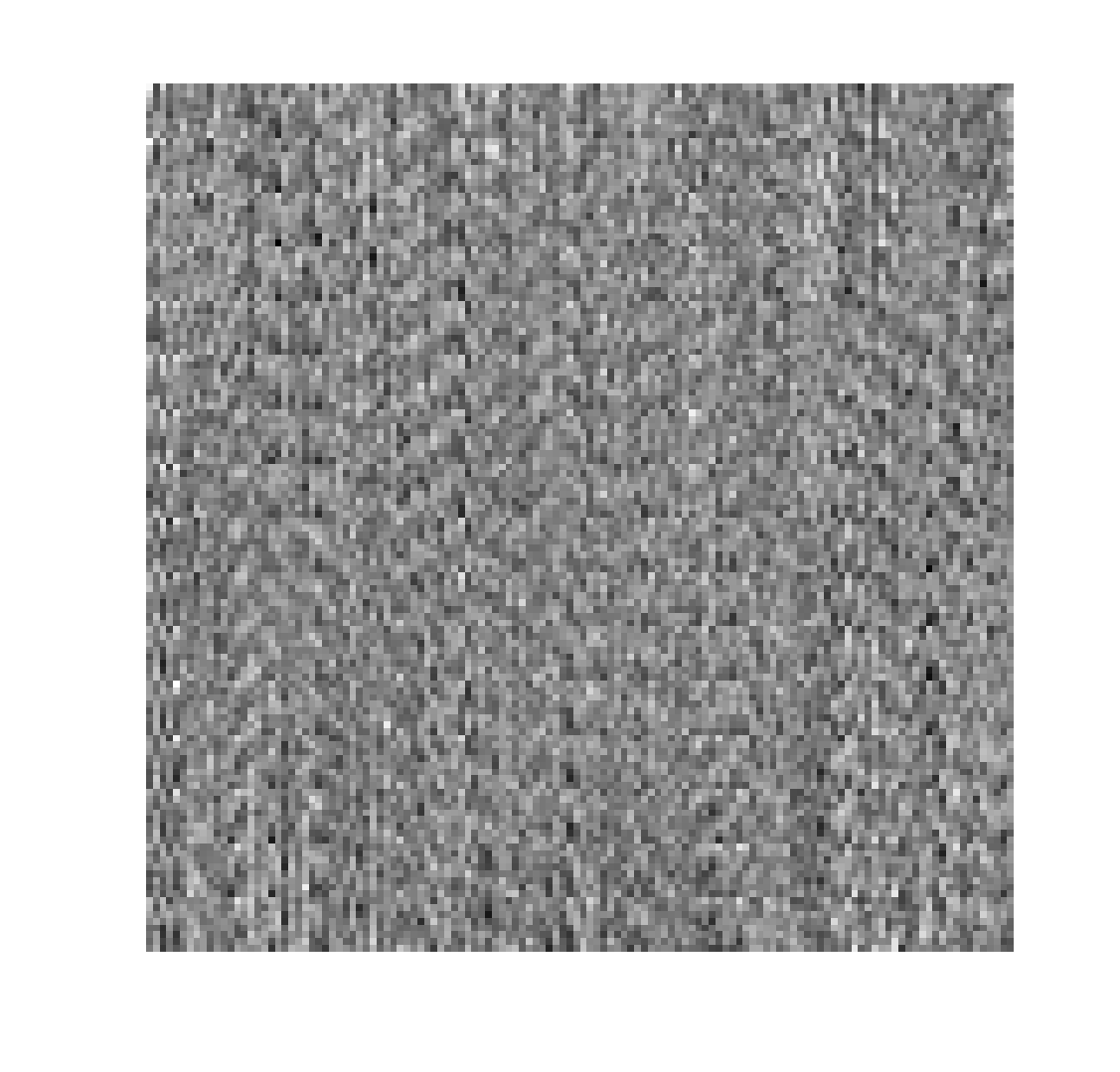,width=4.65cm}
\hspace{-0.6cm}\epsfig{figure=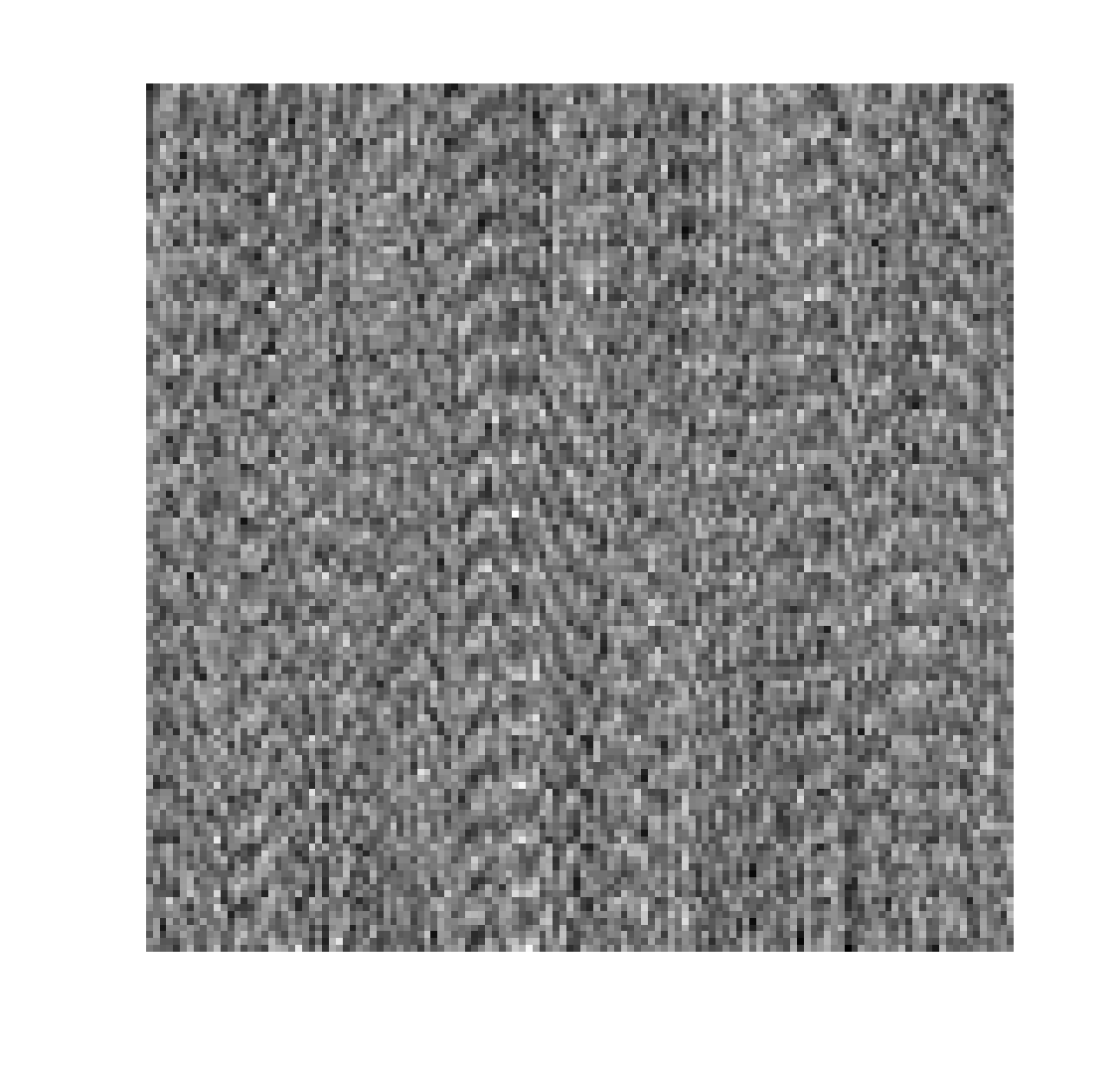,width=4.65cm}}}
\vspace{-0.5cm}
\centerline{(a1) \hspace{3.5cm}
(a2) }
{\centerline{\hspace{0.0cm}\epsfig{figure=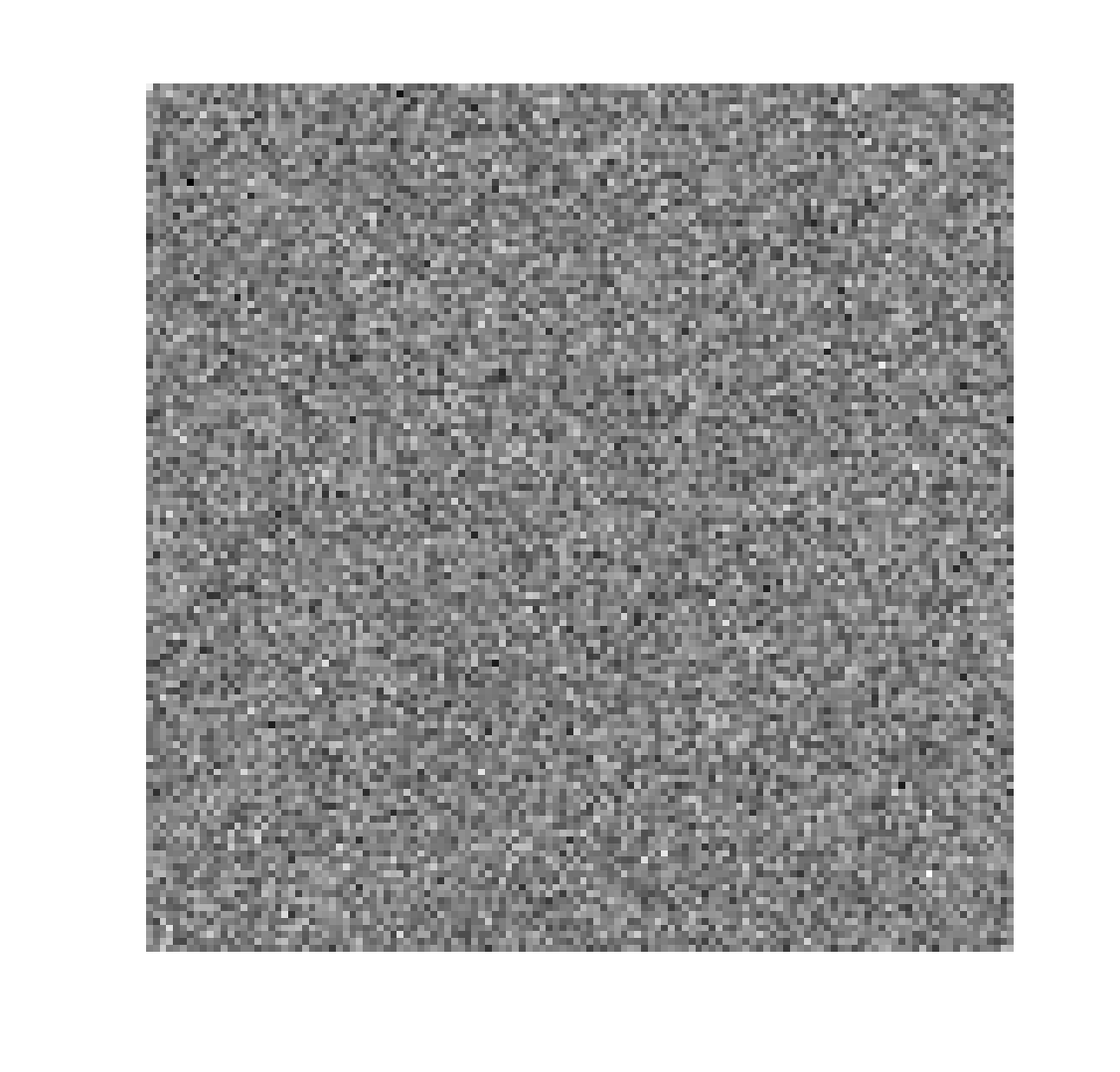,width=4.65cm}
\hspace{-0.6cm}\epsfig{figure=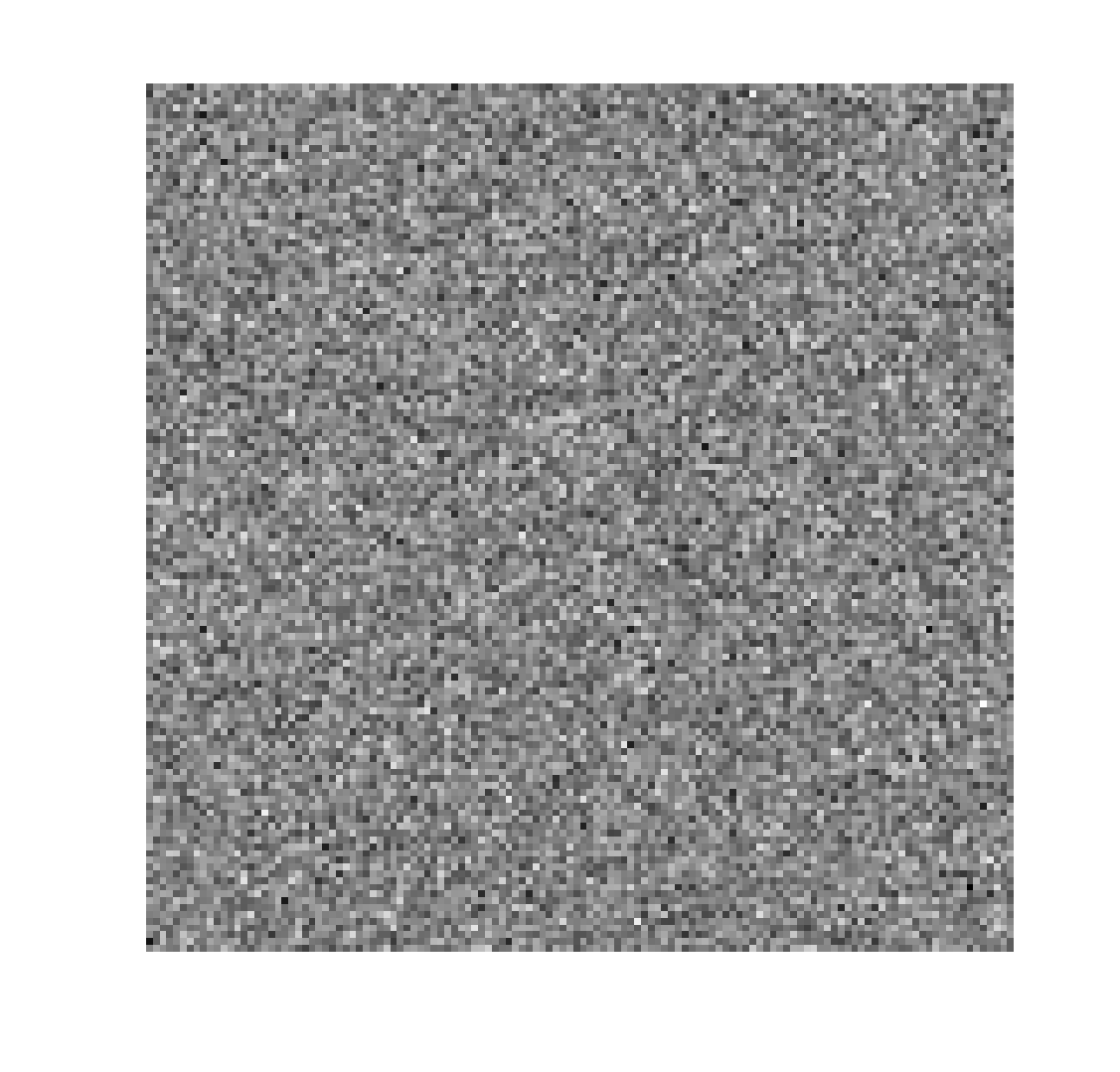,width=4.65cm}}}
\vspace{-0.5cm}
\centerline{(b1) \hspace{3.5cm}
(b2) }
\caption{\label{fig_G}
Matrix $G\in\mathbb{R}^{128\times128}$ of several sparsifying dictionaries $D$ for various $A$. Recall that $D = GAH$ and the sensing matrix of $D$ is $\mathcal E G^{-1}$. 
In (a1) and (b1), $A$ is a Gaussian random matrix. 
In (a2), and (b2), $A$ is a Bernoulli random matrix. 
In (a1) and (a2), $D = D_{KSVD}$.
In (b1) and (b2), $D = D_{wavelet}$.
}
\end{figure}

The entries of $A \in \mathbb R^{l \times n}$ are i.i.d. Bernoulli random variables (where $\pm \frac{1}{\sqrt{n}}$ with equal probability) and Gaussian random numbers (where $\mathcal N(0, n^{-1})$). 
$D$ and $A$ have equal ranks. Experiments were performed on sparsifying dictionaries of $\mathbb R^{128 \times 1024}$, including $D_{wavelet}$ and $D_{KSVD}$, which were derived using the K-SVD algorithm, based on the set of training vectors obtained from the $256 \times 256$ gray-scale test image ``Boat"\footnote{Parameters of the K-SVD process were set at $K=1024$,
$L=64$,
$numIteration=50$,
$InitializationMethod='GivenMatrix'$, and   
$errorGoal=10^{-8}$. 
}. 
The image was divided into overlapping patches of $16 \times 16$ pixels with stride $2$ (in each dimension) and overlaps of $4$ pixels (in each dimension), which resulted in $3,721$ training patches. As the stride is $2$ in each dimension, each patch formed a vector of $128$ pixels. After the mean of each patch was normalized to zero, the resulting patches were transformed into vectors (via the vec operation) to form the set of training vectors. The dictionary
$D_{wavelet}$ was derived from the CDF 9-7 wavelet. 
The first $640$ columns (i.e., each level has $128$ columns and there are $5$ levels) in $D_{wavelet}$ were obtained from the first $5$ levels (the support of a wavelet at level $6$ is larger than the size of an image patch), whereas the remainder were generated randomly. The column norms of the sparsifying dictionaries were normalized. Figures~\ref{fig_D} and \ref{fig_G} present the sparsifying dictionaries and corresponding matrices $G$, derived for various $A$ in accordance with (\ref{rangeConstructions}). Note that $G$ is related to the sensing matrix of $D$.

\begin{figure*}[tp]
\centering
{\epsfig{figure=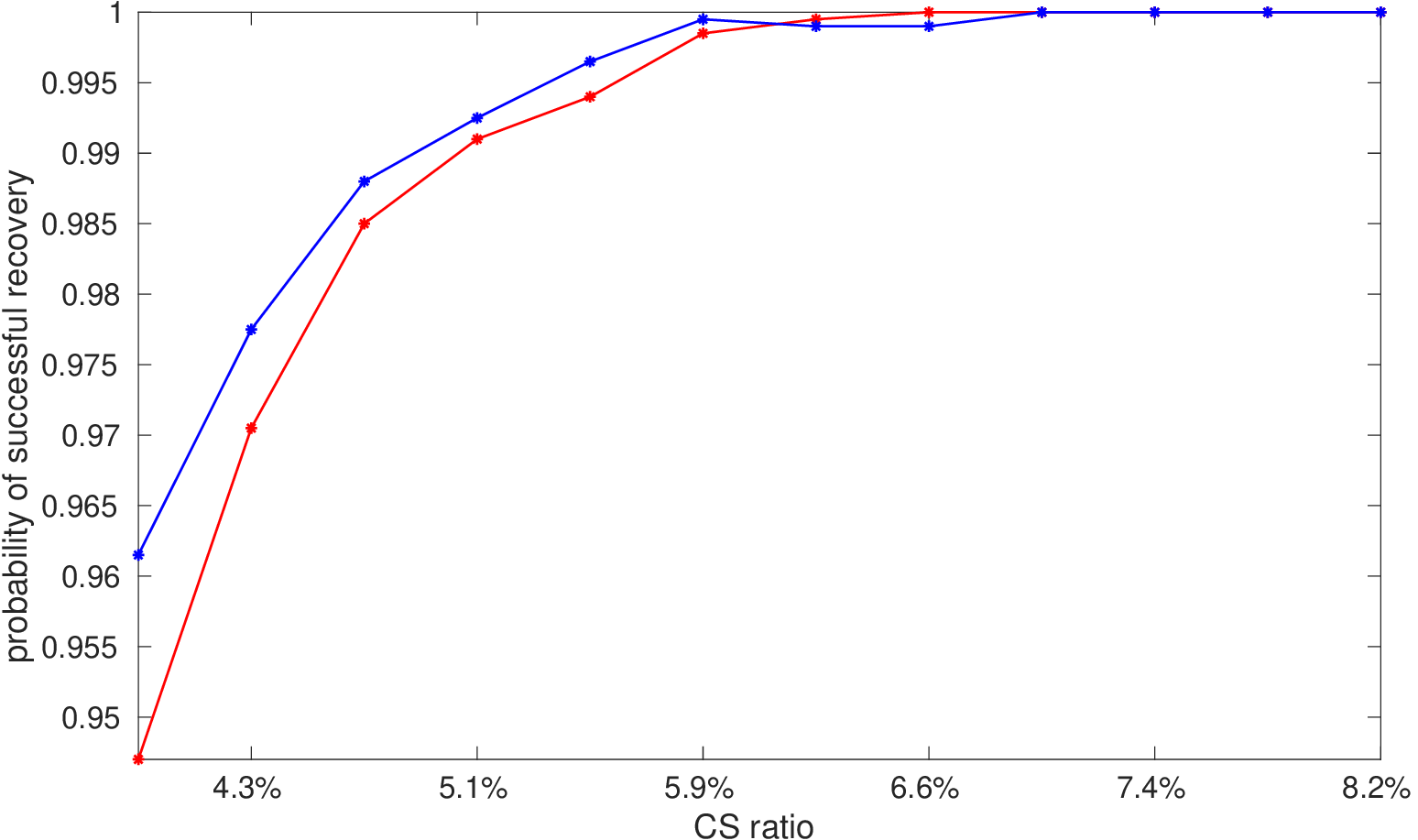, width=8.5cm}}
  \hspace{0.4cm}  
  {\epsfig{figure=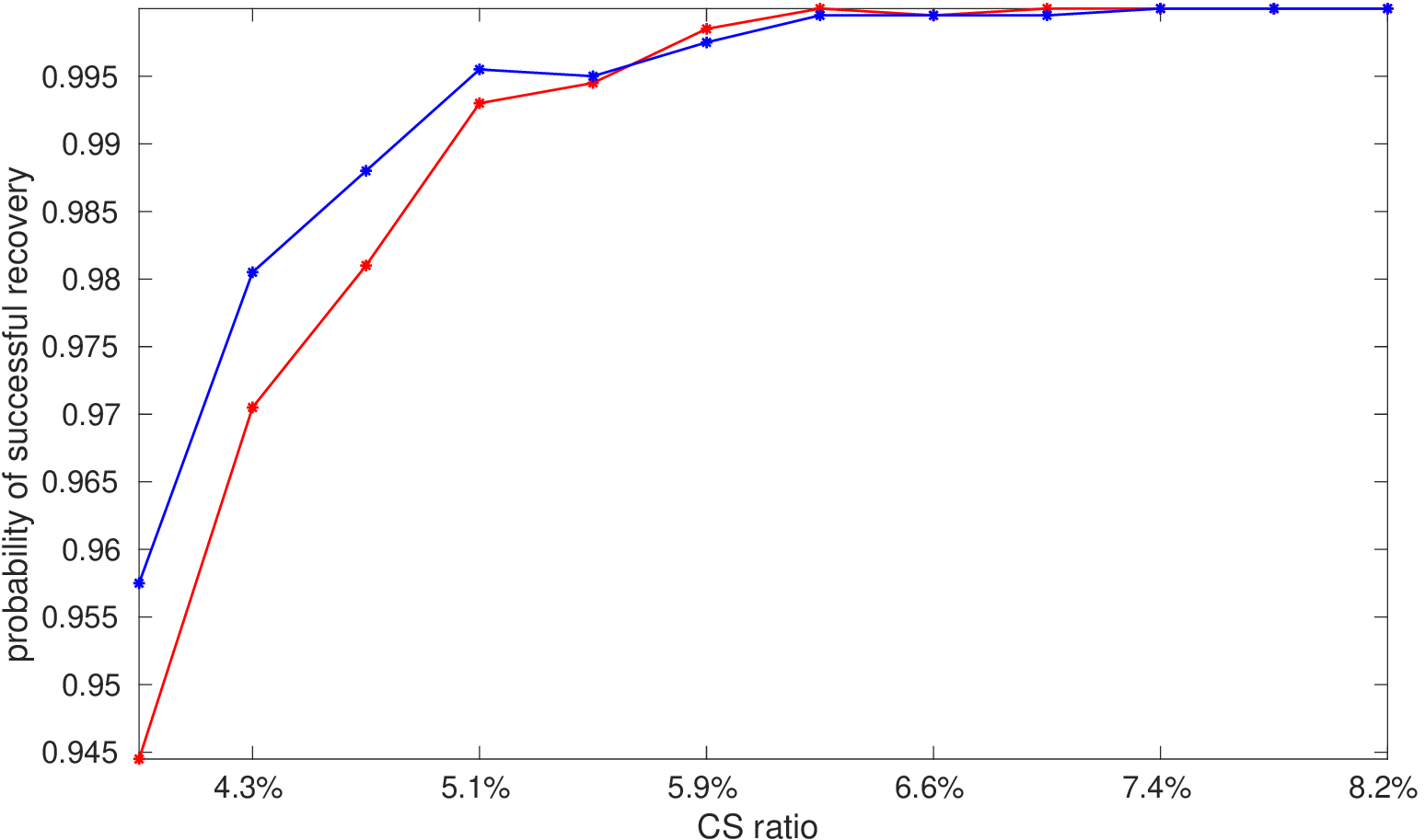, width=8.5cm}} 
  \vspace{0.2in}
\centerline{(a1) $k = 10$; $A$ is Gaussian \hspace{4cm} (a2): $k=10$; $A$ is Bernoulli}
{\epsfig{figure=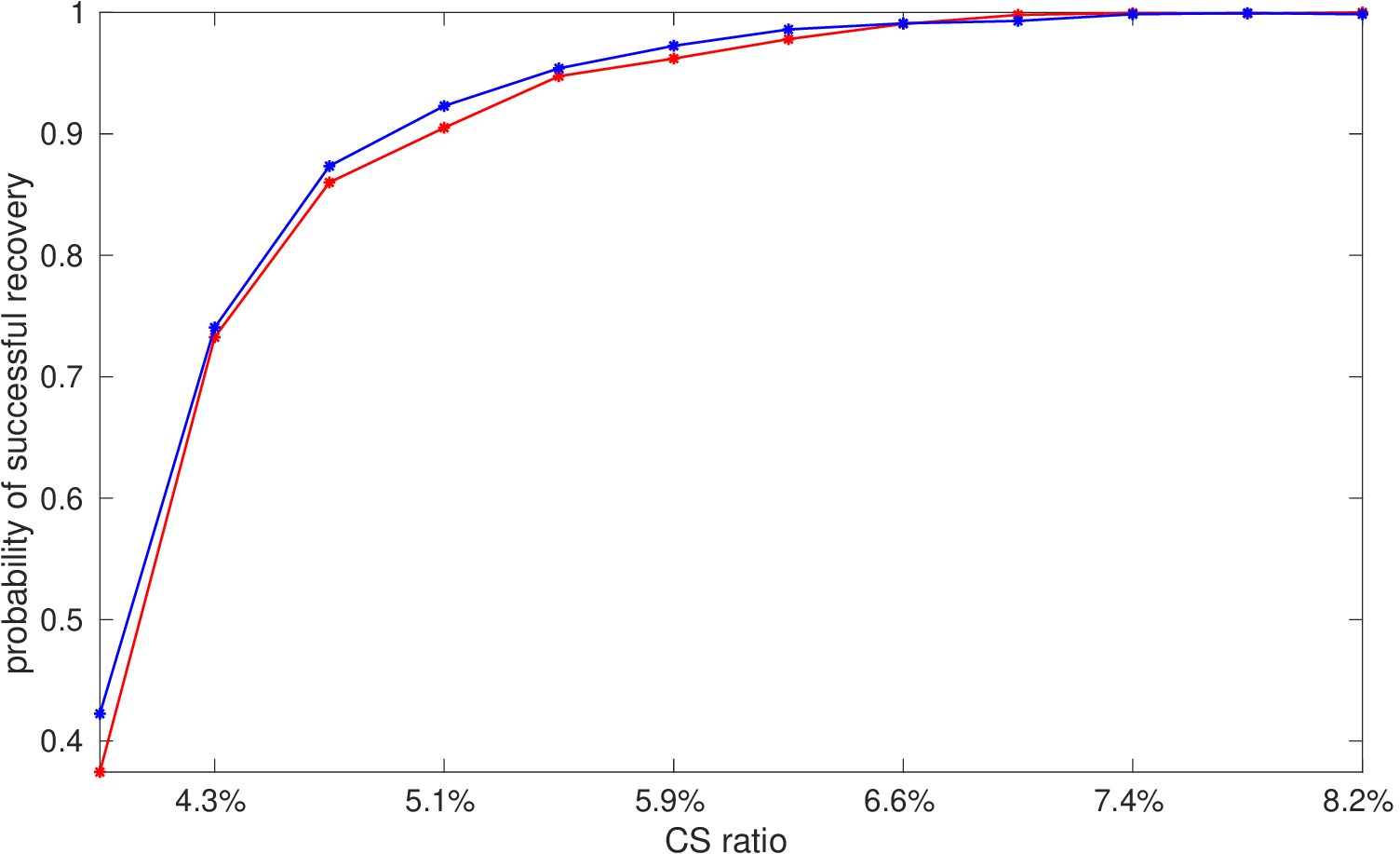, width=8.5cm}}
  \hspace{0.4cm}  
  {\epsfig{figure=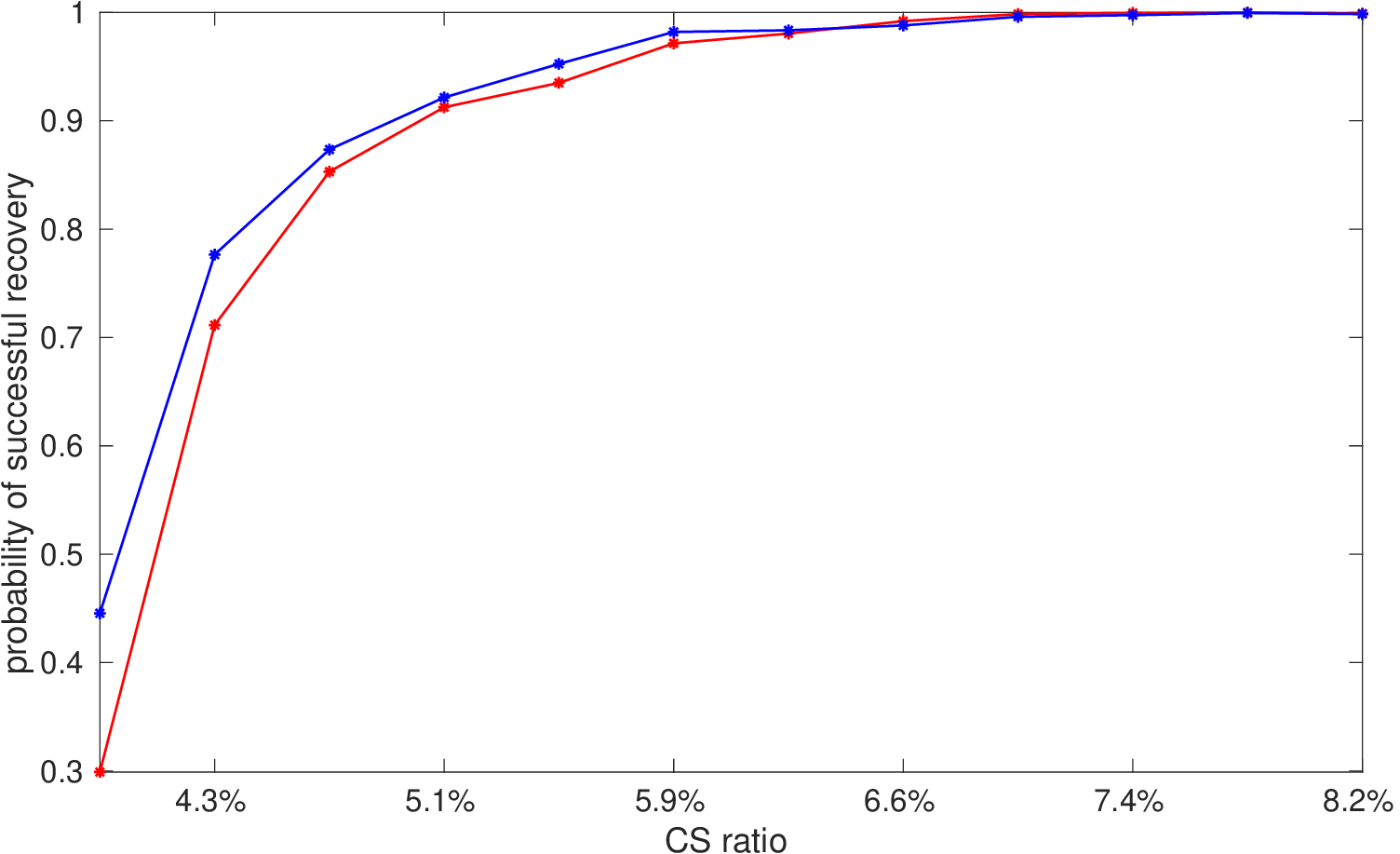, width=8.5cm}}
    \vspace{0.2in}
\centerline{(b1) $k = 12$; $A$ is Gaussian \hspace{4cm} (b2): $k=12$; $A$ is Bernoulli}
\centering
{\epsfig{figure=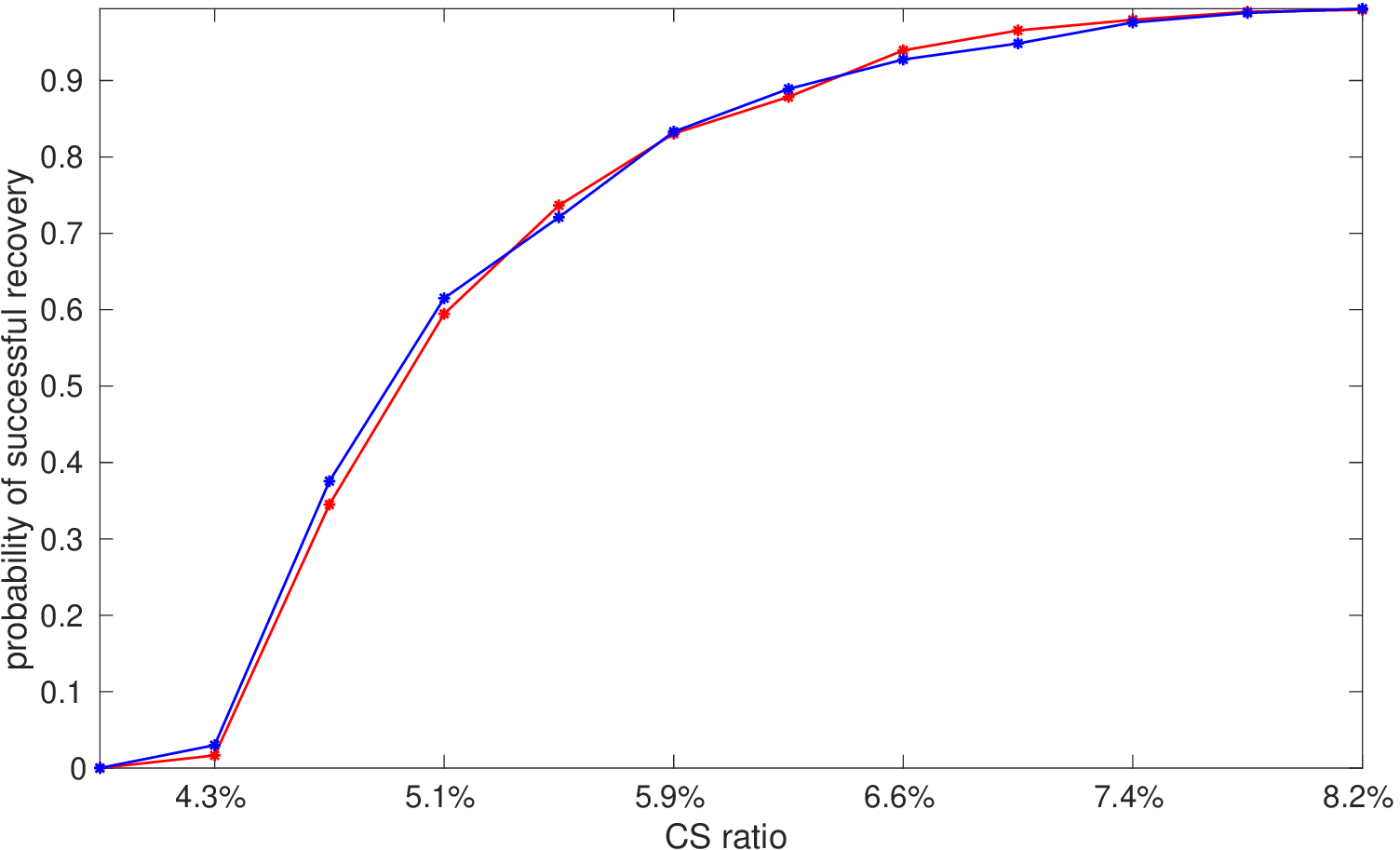, width=8.5cm}}
  \hspace{0.4cm}  
  {\epsfig{figure=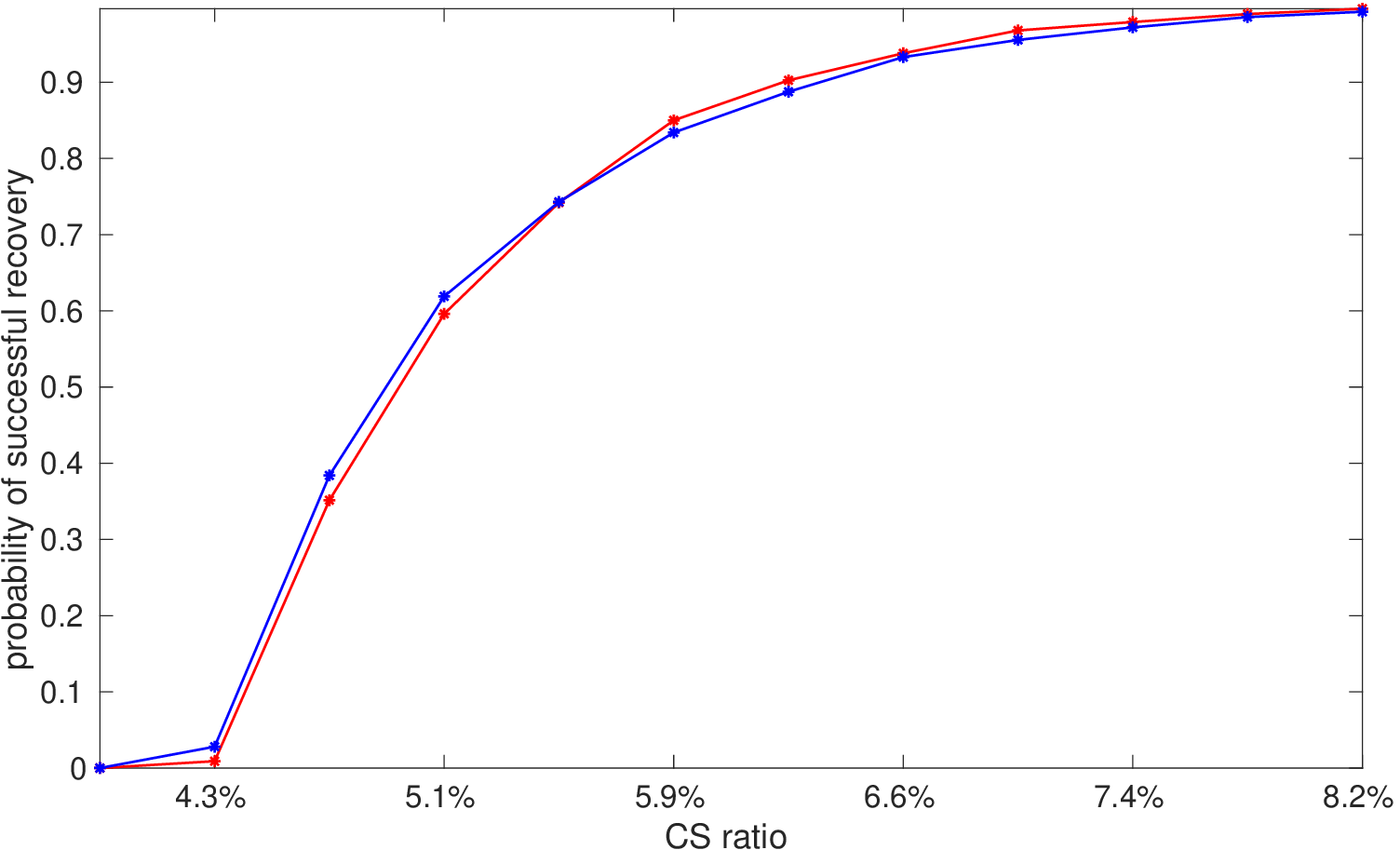, width=8.5cm}}
    \vspace{0.2in}
\centerline{(c1) $k = 14$; $A$ is Gaussian \hspace{4cm} (c2): $k=14$; $A$ is Bernoulli}
  
\caption{\label{fig_KSVD}
Comparisons of CS recovery performance (i.e., the probability of sparse vector recovery versus CS ratio) using  sparsifying dictionary $D_{KSVD}$. Red and blue curves were respectively obtained using the benchmark (\ref{cs_p2}) and our approach (\ref{cs_p1}). Sparse vectors $x$ were randomly generated and each point on the curve is the average of $2,000$ probability measurements. The positions of non-zero coefficients of $x$ are uniformly distributed and the values of the non-zero coefficients of $x$ are uniformly distributed in $[-1, 1]$. 
In (a1) and (a2), sparsity level is $10$; in (b1) and (b2), sparsity level is $12$; and in (c1) and (c2), sparsity level is $14$. }
\end{figure*}

\begin{figure*}[tp]
\centering
{\epsfig{figure=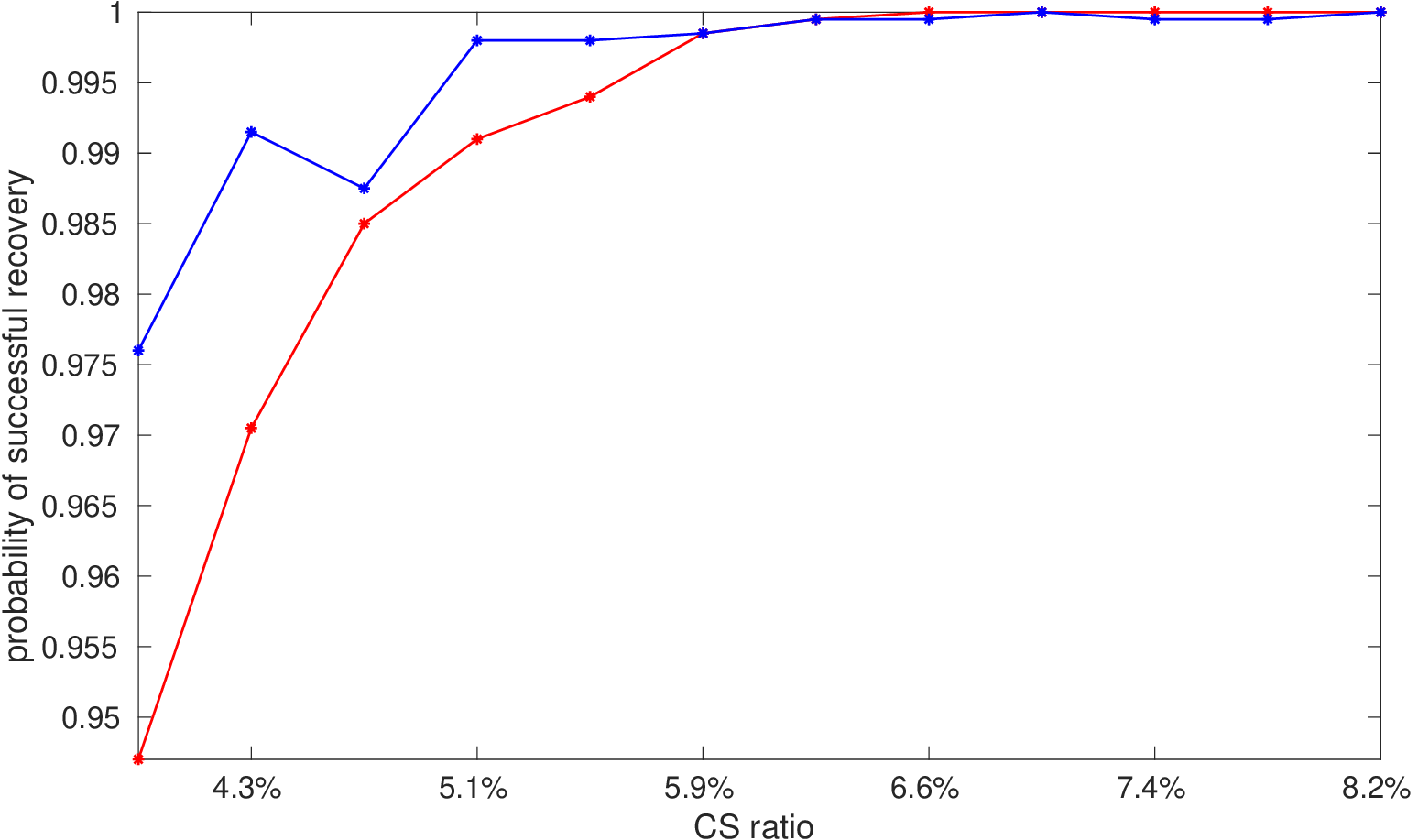, width=8.5cm}}
  \hspace{0.4cm}  
  {\epsfig{figure=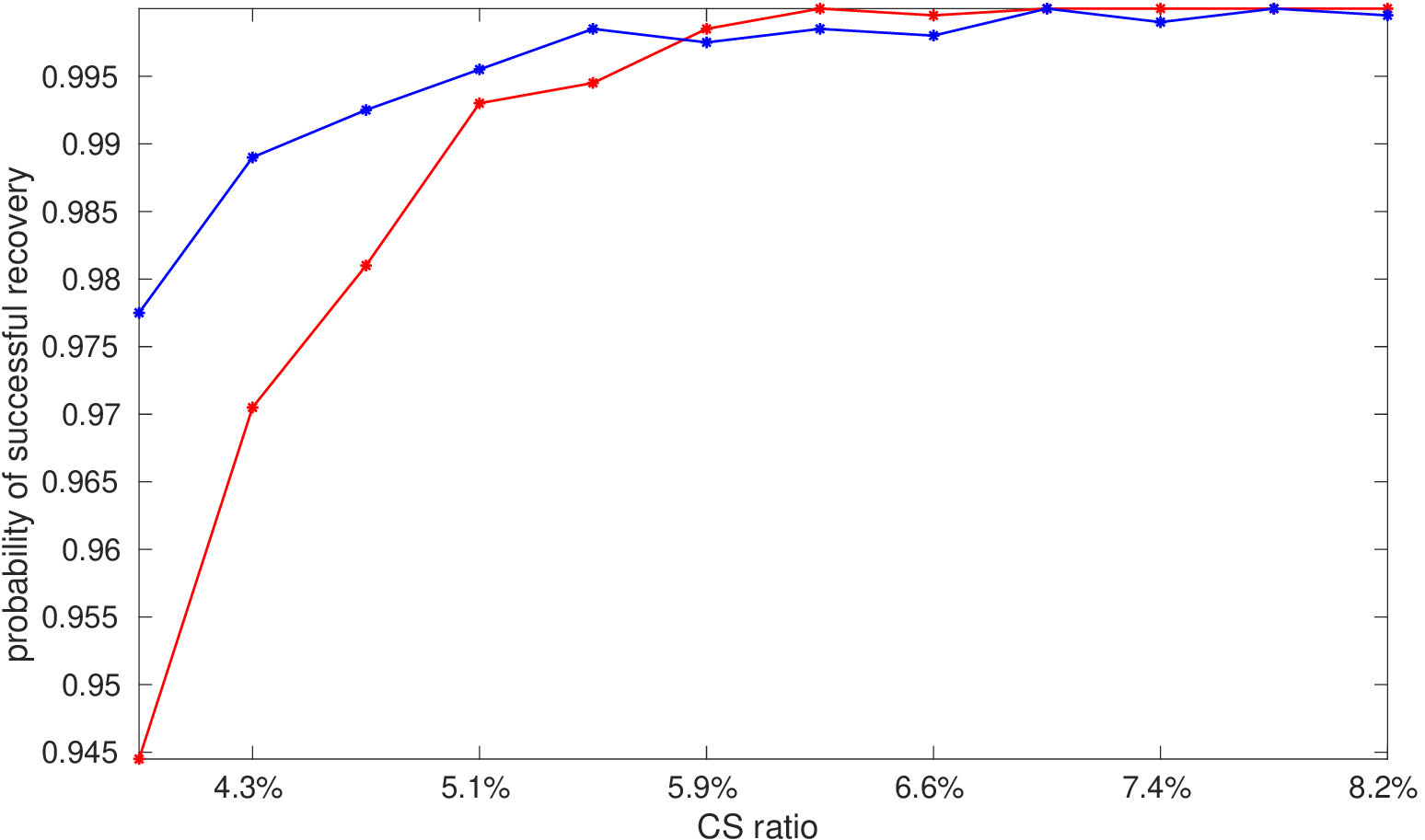, width=8.5cm}}
  \vspace{0.2in}
\centerline{(a1) $k = 10$; $A$ is Gaussian \hspace{4cm} (a2): $k=10$; $A$ is Bernoulli}
\centering
{\epsfig{figure=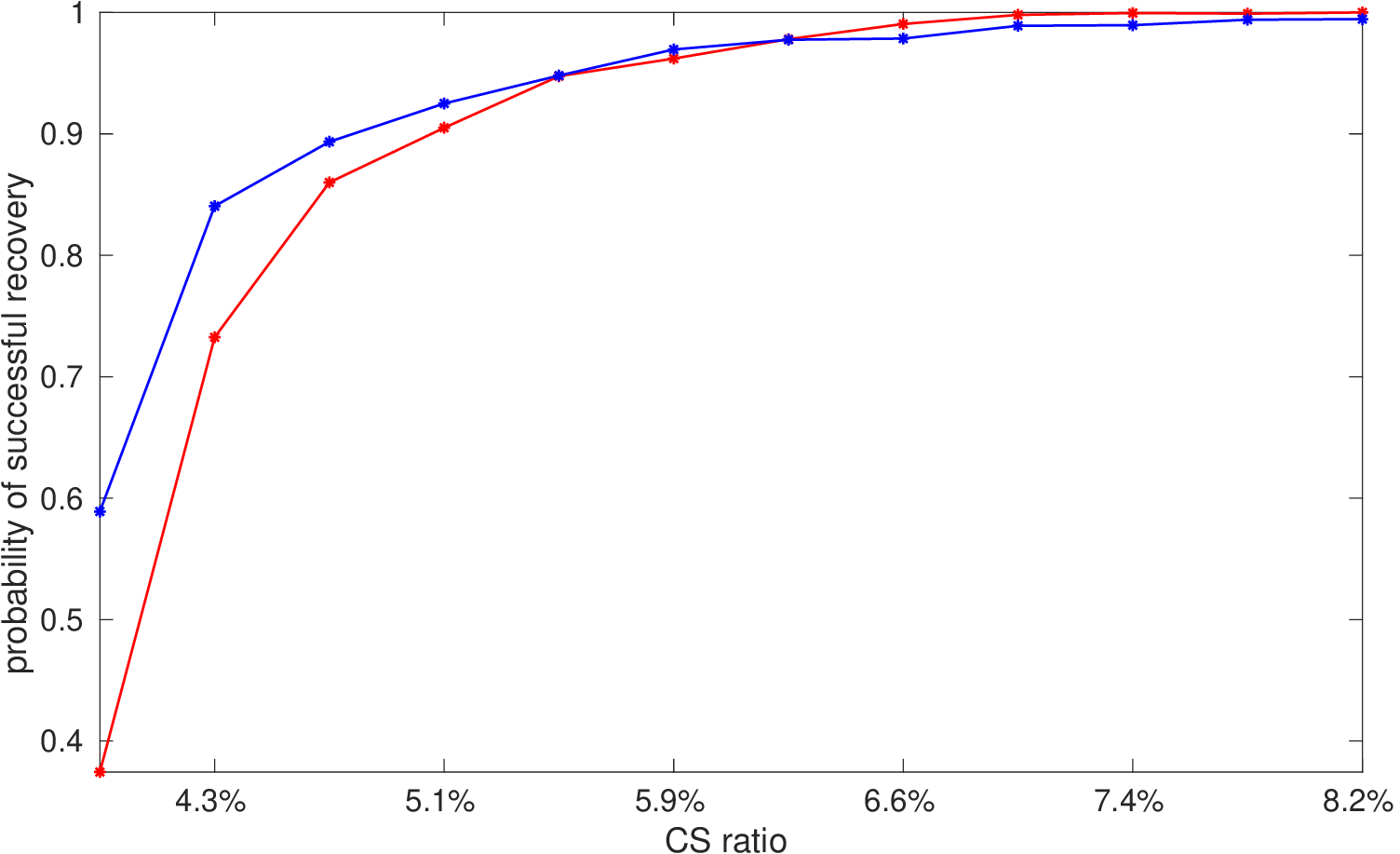, width=8.5cm}}
  \hspace{0.4cm}  
  {\epsfig{figure=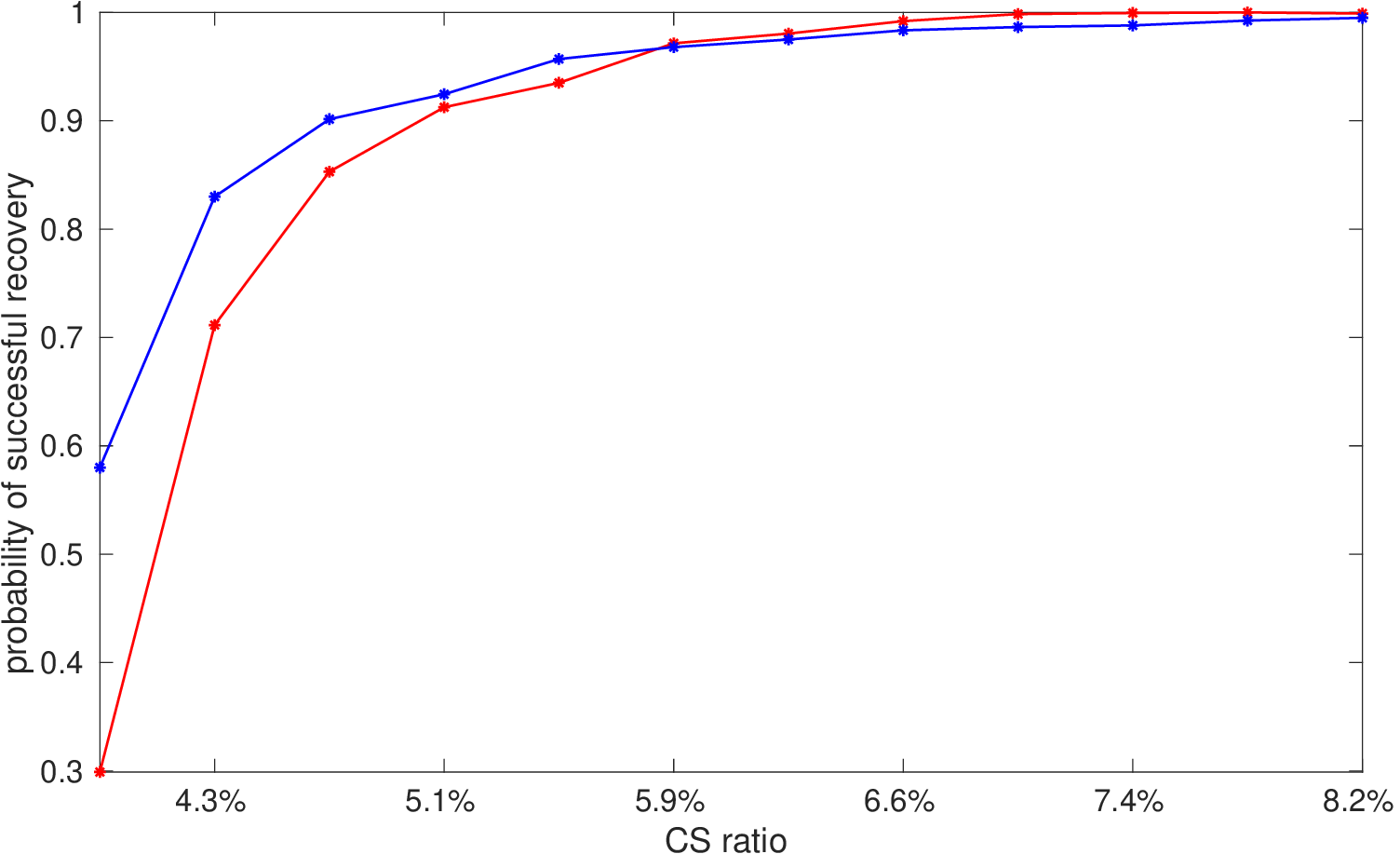, width=8.5cm}}
  \vspace{0.2in}
\centerline{(b1) $k = 12$; $A$ is Gaussian \hspace{4cm} (b2): $k=12$; $A$ is Bernoulli}
\centering
{\epsfig{figure=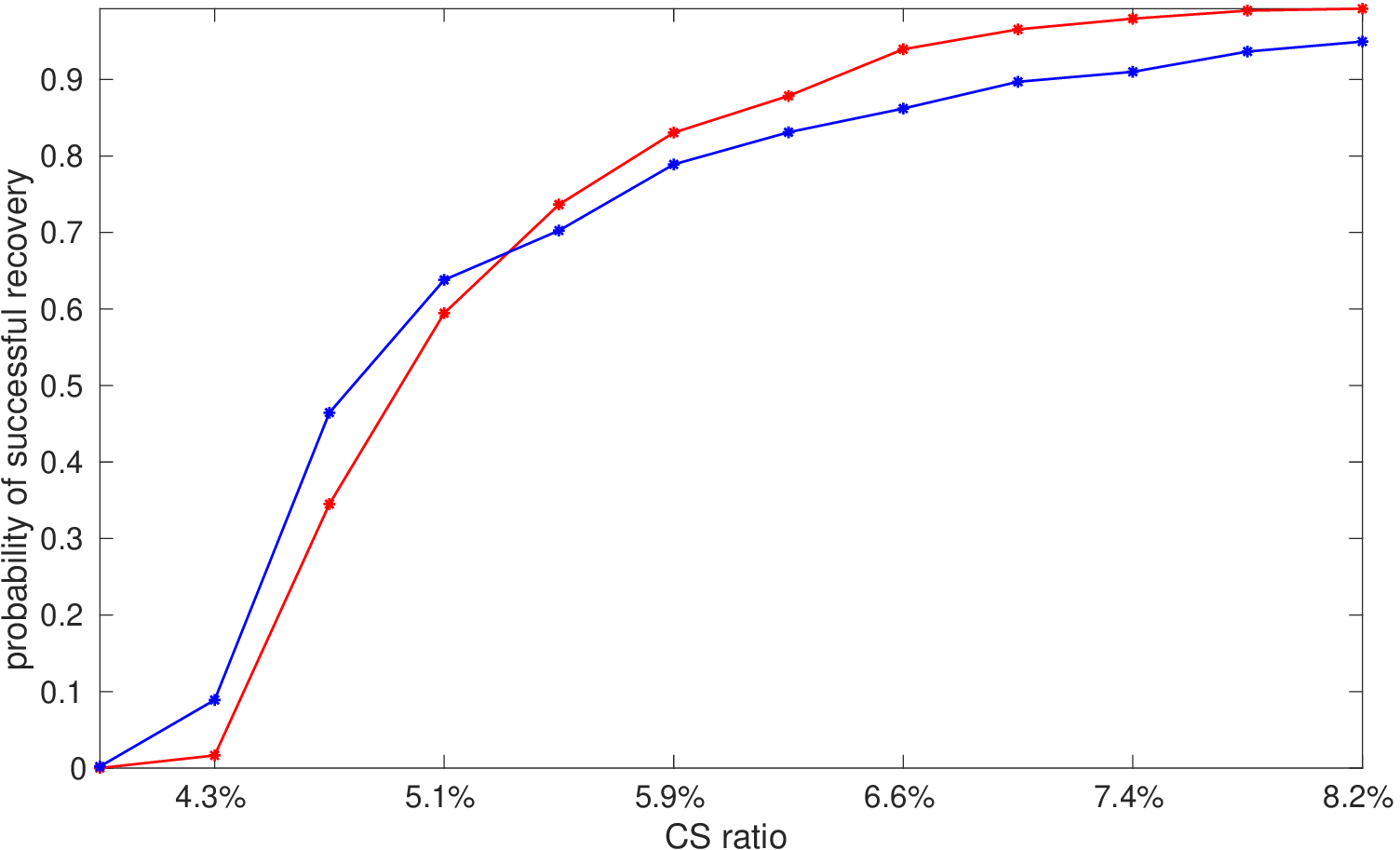, width=8.5cm}}
  \hspace{0.4cm}  
  {\epsfig{figure=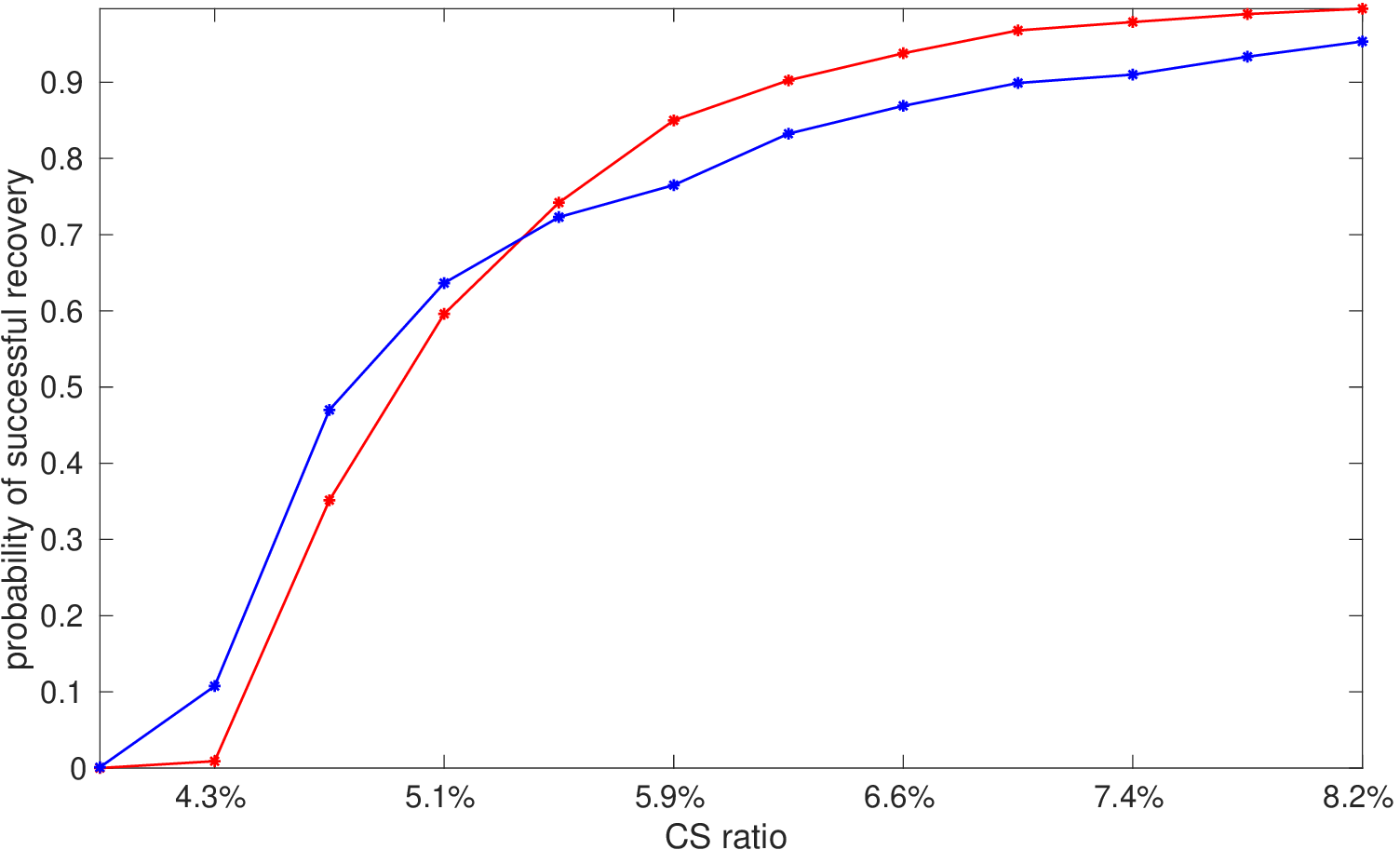, width=8.5cm}}
\centerline{(c1) $k = 14$; $A$ is Gaussian \hspace{4cm} (c2): $k=14$; $A$ is Bernoulli}
  
\caption{\label{fig_wavelet}
Comparisons of CS recovery performance (i.e., the probability of sparse vector recovery versus CS ratio) using  sparsifying dictionary $D_{wavelet}$ (CDF 9/7). Red and blue curves were respectively obtained using the benchmark (\ref{cs_p2}) and our approach (\ref{cs_p1}). Sparse vectors $x$ were randomly generated and each point on the curve is the average of $2,000$ probability measurements. The positions of non-zero coefficients of $x$ are uniformly distributed and the values of the non-zero coefficients of $x$ are uniformly distributed in $[-1, 1]$.
In (a1) and (a2), sparsity level is $10$; in (b1) and (b2), sparsity level is $12$; and in (c1) and (c2), sparsity level is $14$. }
\end{figure*}

Under Theorem~\ref{Thm:Main}, the following two optimizations derive the same solution, if $\mathcal E G^{-1}D = \mathcal AH$:
\begin{align}
\begin{cases}
\displaystyle \min_{x}  \| z - \mathcal E G^{-1} D x \|^2_2 \\
\| x  \|_0 \leq k, \label{cs_p1}
\end{cases}
\end{align}
and 
\begin{align}
\begin{cases}
\displaystyle \min_{x}  \| z- \mathcal E A x \|^2_2 \\
\| x \|_0 \leq k  \label{cs_p2}.
\end{cases}
\end{align}
Note that $z$ in (\ref{cs_p1}) and (\ref{cs_p2}) are low-dimensional observations of $x$ obtained from $\mathcal E A x$ and $\mathcal E$ in (\ref{cs_p1}) and (\ref{cs_p2}) are the same.
To verify that (\ref{cs_p1}) yields performance similar (in terms of probability) to the the sparse recovery obtained using (\ref{cs_p2}), experiments were performed on the sparsifying dictionaries ($D_{wavelet}$ and $D_{KSVD}$) and using the compressive sampling matched pursuit (CoSaMP) algorithm \cite{needell2009cosamp} for the recovery of sparse vectors. Figures \ref{fig_KSVD} and \ref{fig_wavelet} illustrate CS recovery performance using plots indicating the probability of successfully recovering a sparse vector versus the CS ratio. The figures compare the performance of our approach (\ref{cs_p1}) versus the benchmark (\ref{cs_p2}) using Gaussian and Bernoulli sensing matrices at various levels of sparsity. 
The horizontal and vertical axes respectively indicate the CS ratio (i.e.,  $\frac{m}{n}\times 100 \%$, where $m$ is the number of rows of $\mathcal E$ and $n=1024$) and the probability of successfully recovering a sparse vector.
We claim that the true sparse vector $x$ can be recovered as long as estimate $\hat x$ satisfies $\| \hat x - x \|_1  < 1024 \times 10^{-2}$. 

\subsection{MRI image recovery}

Significant research effort based on the sparse representation has been directed towards finding ways to accelerate MR imaging. At the heart of these sparse reconstructions assumes MR images can be under-sampled in such a way that data collection time can be dramatically reduced while maintaining image quality. 
Here we report experimental results that demonstrate our proposed MRI recovery algorithm on data from the publicly available fastMRI dataset \cite{zbontar2018fastmri}. Our aim is not to develop an algorithm to achieve the state-of-the-art performance for MRI image recovery (baseline deep neural network methods such as variational networks and U-nets~\cite{fastMRI2019,2020fastMRIChallenge}). The experiment is designed to demonstrate that using the proposed matrix factorization method can be a promising approach for recovering under-sampled MRI images. In this application one of the factors in our proposed decomposition is adapted to the acceleration mask (restricted by physically prescribed sensing mechanism), which otherwise does not allow for direct RIP based recovery guarantees in the $\ell_1$-synthesis approach to CS. A recent survey of applying the compressed sensing technique to the MRI image recovery \cite{ye2019compressed} presents a variety of techniques to improve the quality of recovered images.  In this study, we compared the performance of our method to that based on the TV approach, which promotes sparsity of spacial gradients by approximates a solution to (see~\cite{Chambolle04,landi2008total,panic2020mri})
\begin{align}\label{TValgorithm}
\argmin_{Z\in\R^{n_1\times n_2}} 
\|Y - \mathcal{R} F_1 Z F_2 \|_F^2
+ \lambda\sum_{i,j}  \|\nabla Z (i,j) \|_2.
\end{align}

\begin{figure}[h!]
\begin{center}
\epsfig{figure=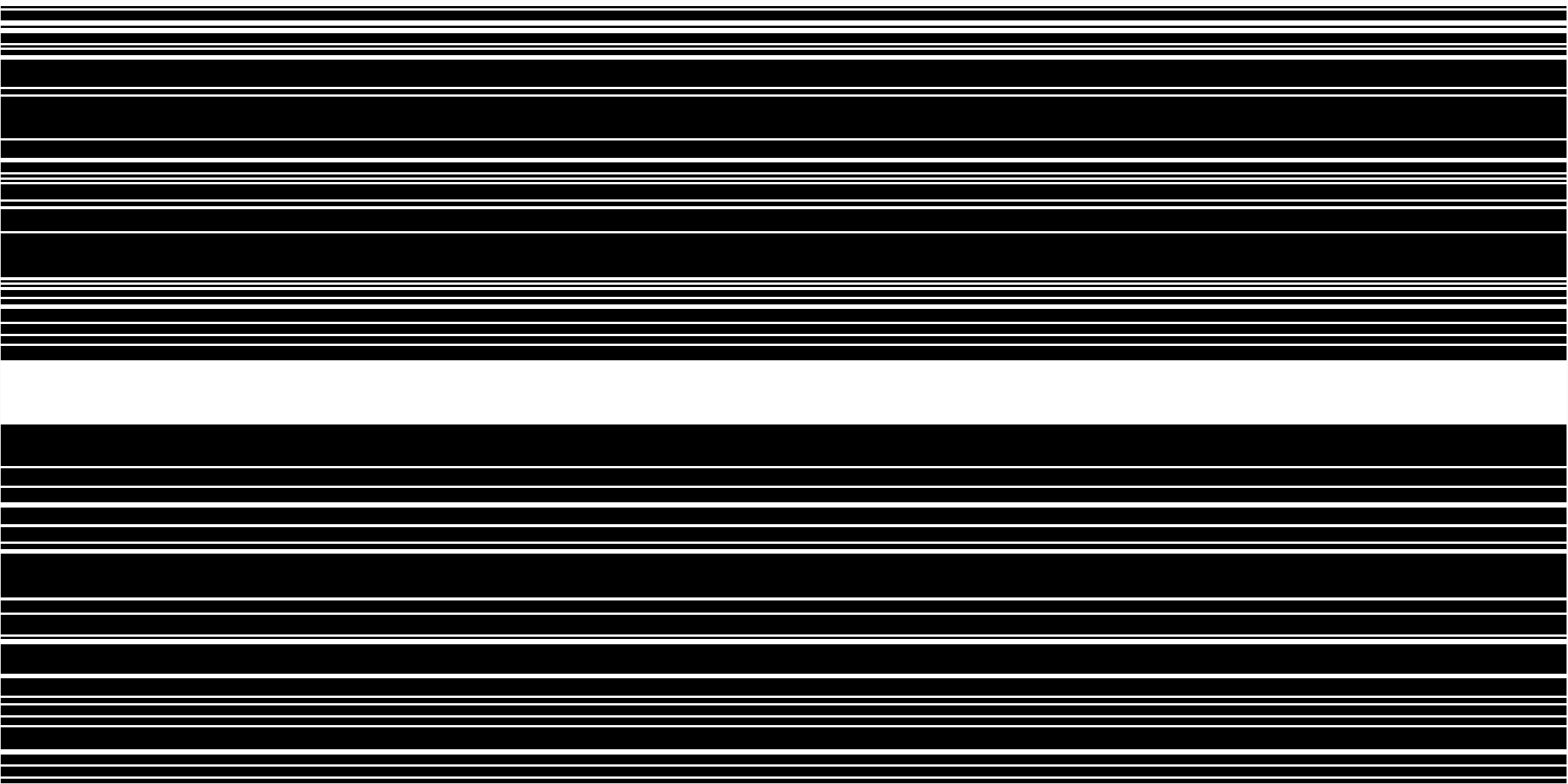,width=0.48\columnwidth,trim = 0cm 0cm 0cm 0cm,clip}
\epsfig{figure=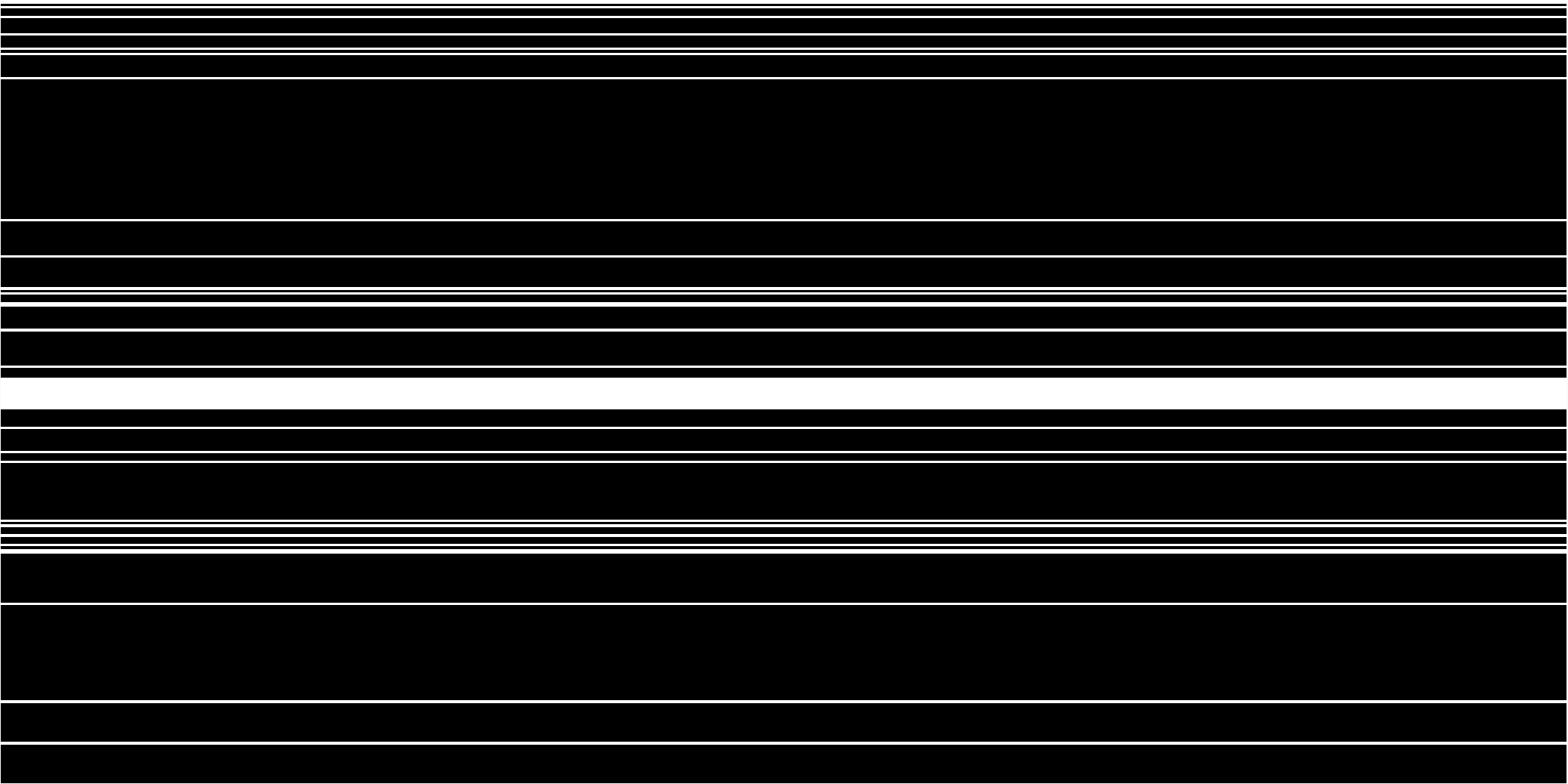,width=0.48\columnwidth,trim = 0cm 0cm 0cm 0cm,clip}
\caption{
Scan line locations for 4-fold (left) and 
8-fold (right) undersampling, including fully sampled low-frequency region. 
}
\label{fig_mask}
\vspace{-0.5cm}
\end{center}
\end{figure}

Under-sampling masks (restrained by the instrumental conditions) for 4-fold and 8-fold acceleration were applied to full $k$-space data according to \eqref{undersampling}. The masks sample 25\%, resp. 12.5\%, of horizontal scan lines including a fully sampled center region - corresponding to low spatial frequencies - containing 8\%, resp. 4\%, of scan lines, as shown in  Fig.~\ref{fig_mask}.
As sparsifying dictionaries we chose CDF 9/7 wavelets (3-levels).

\begin{figure}[t!]
\centering
\includegraphics[width=0.3333333\columnwidth,trim = 0cm 0cm 0cm 0cm,clip]{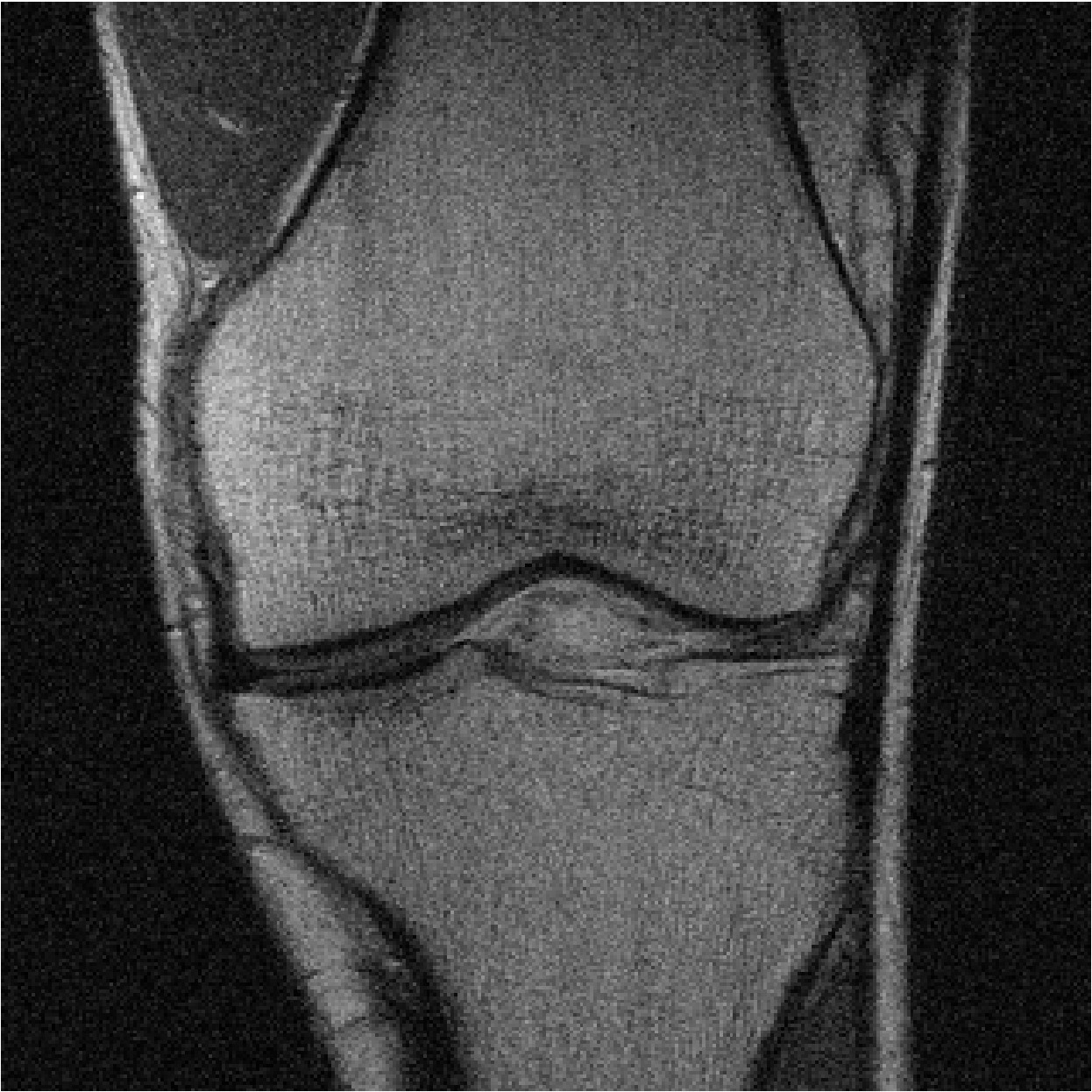}\hfill
\includegraphics[width=0.3333333\columnwidth,trim = 0cm 0cm 0cm 0cm,clip]{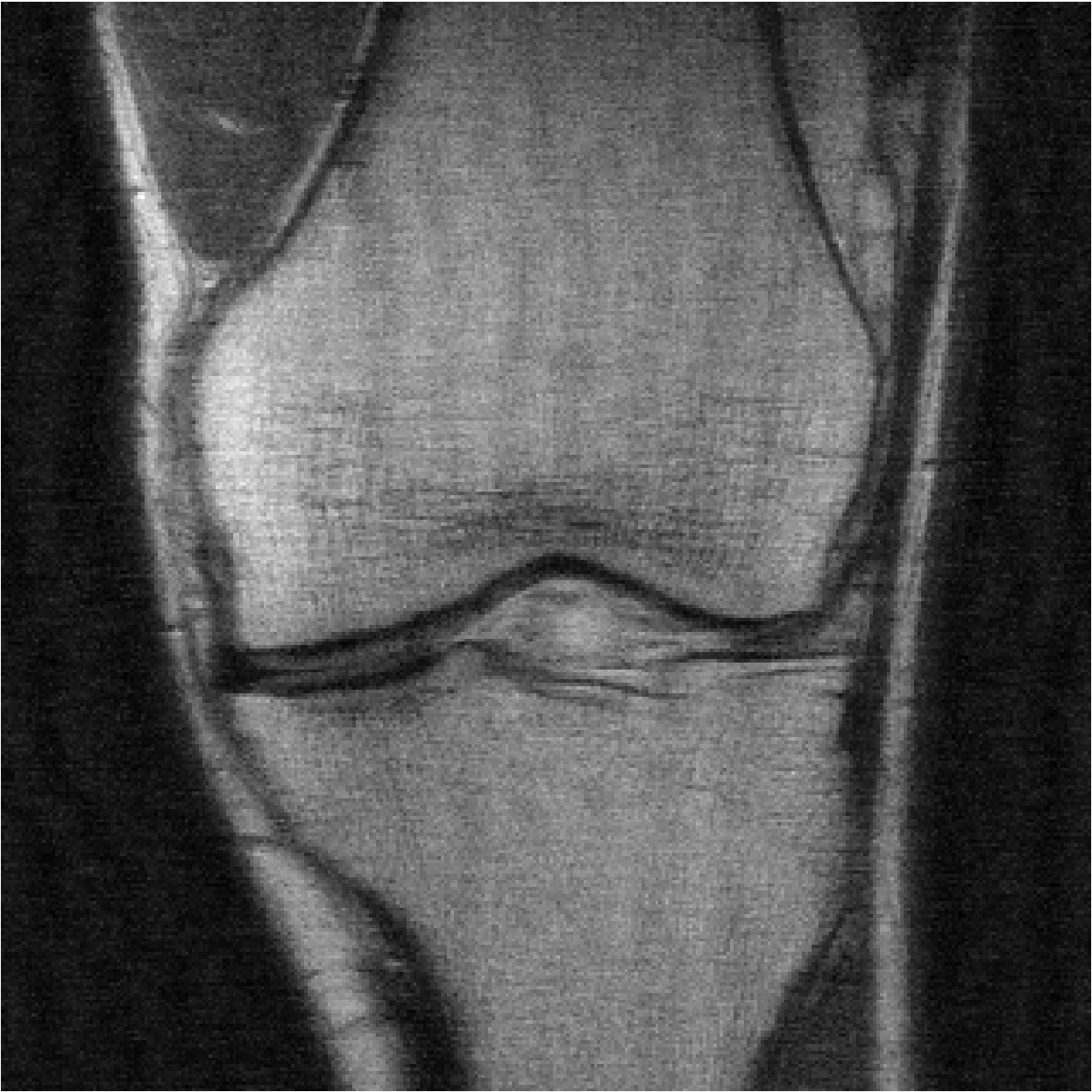}\hfill
\includegraphics[width=0.3333333\columnwidth,trim = 0cm 0cm 0cm 0cm,clip]{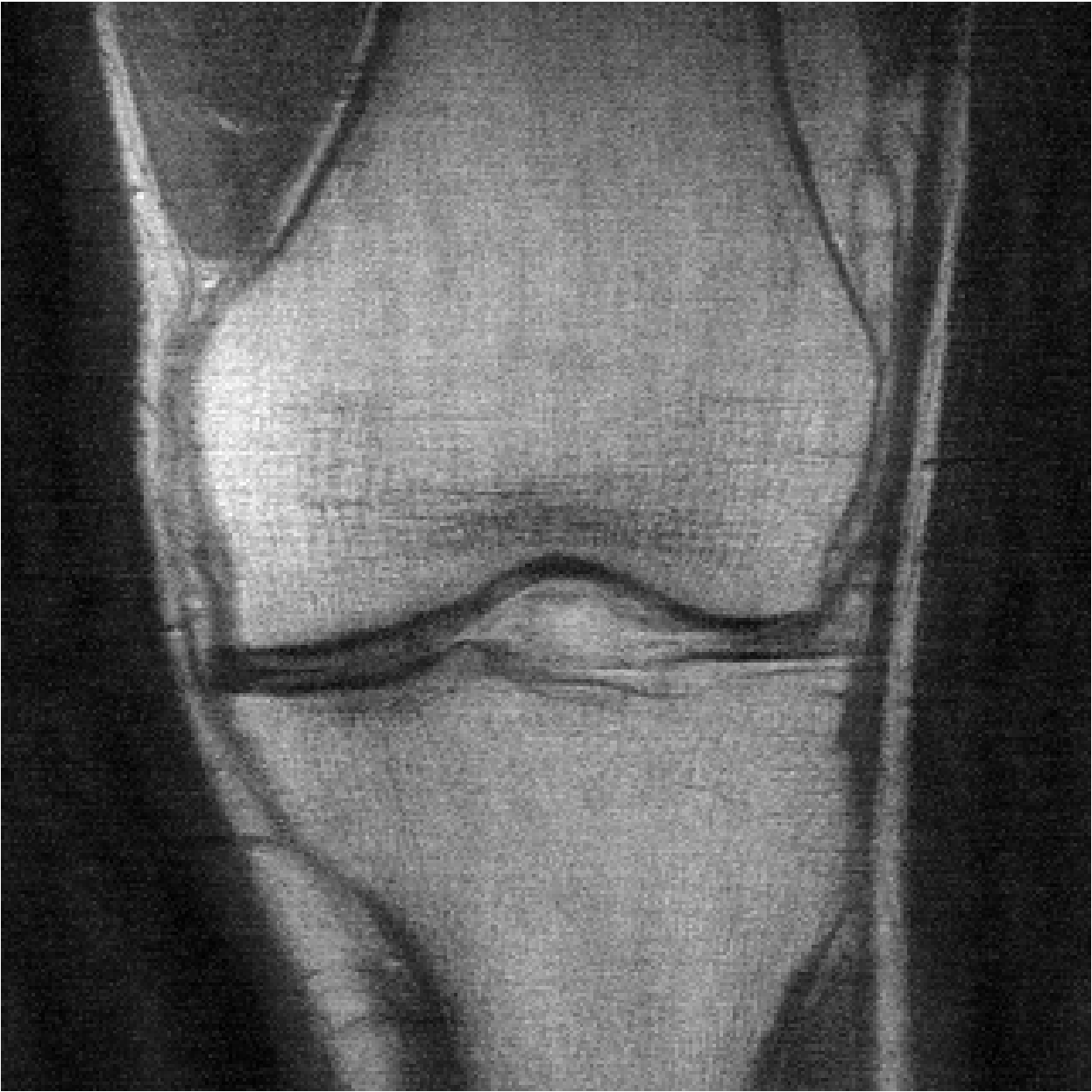}\\
\hspace{2.82cm}
\includegraphics[width=0.3333333\columnwidth,trim = 0cm 0cm 0cm 0cm,clip]{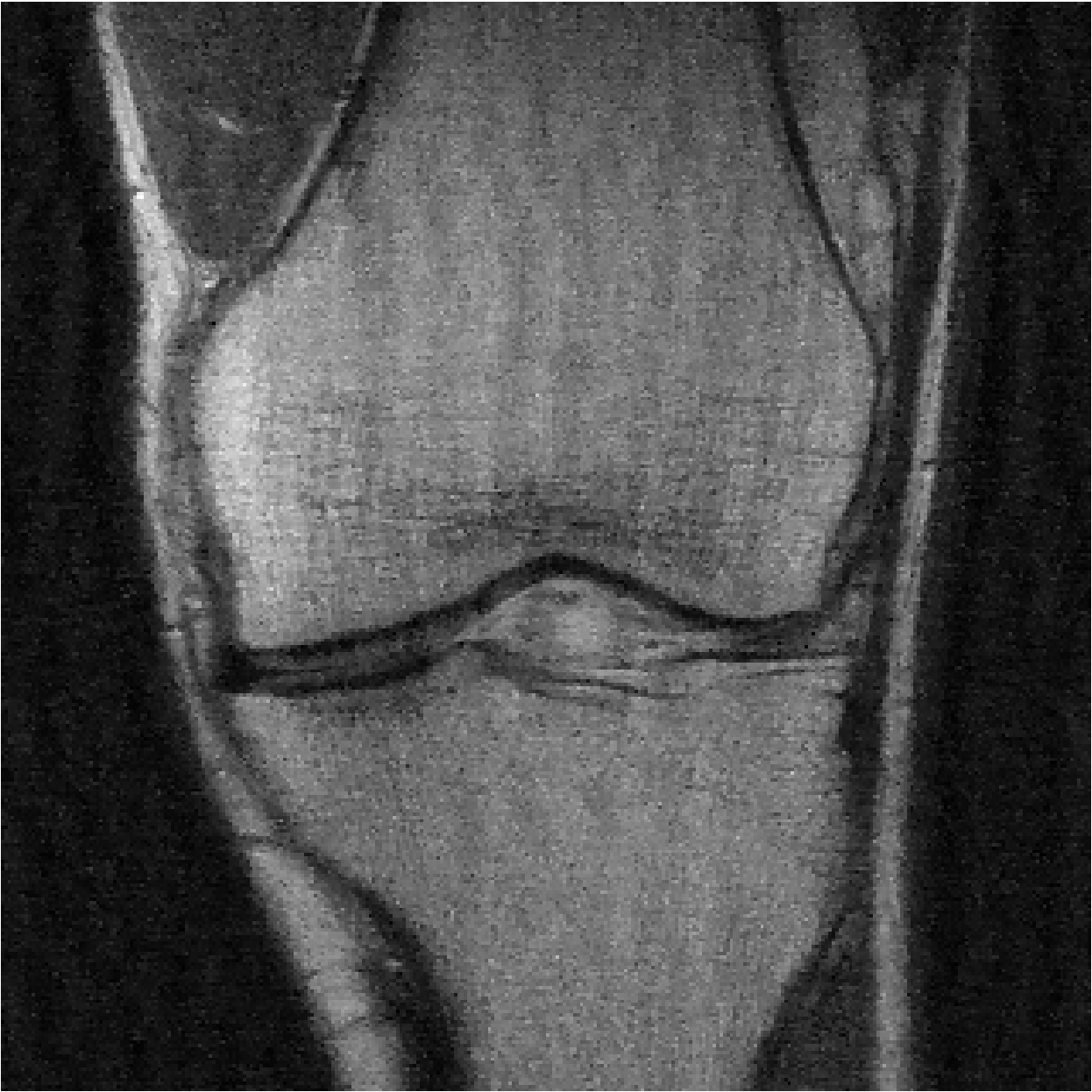}\hfill
\includegraphics[width=0.3333333\columnwidth,trim = 0cm 0cm 0cm 0cm,clip]{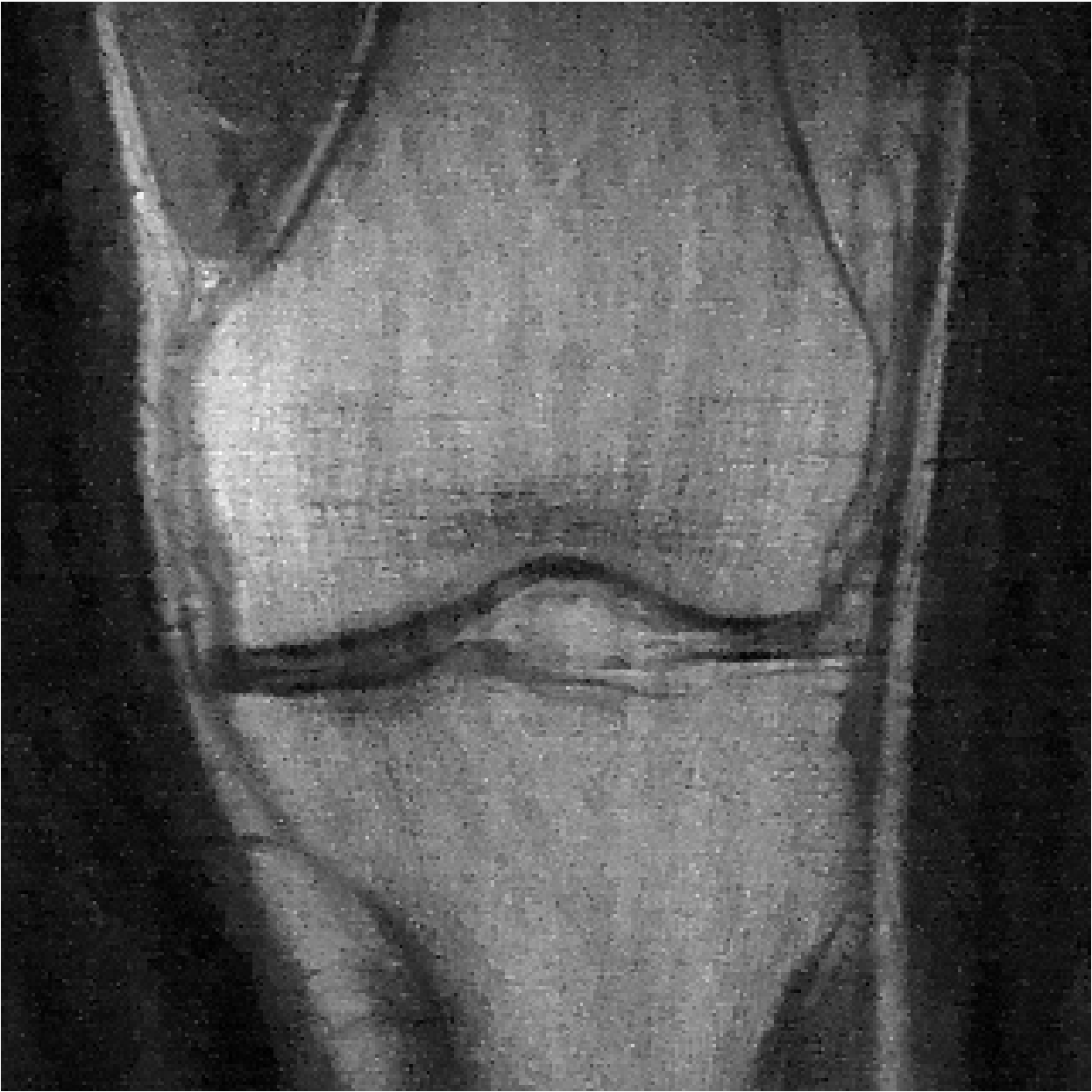}\\
\vspace{0.1cm}
\includegraphics[width=0.3333333\columnwidth,trim = 0cm 0cm 0cm 0cm,clip]{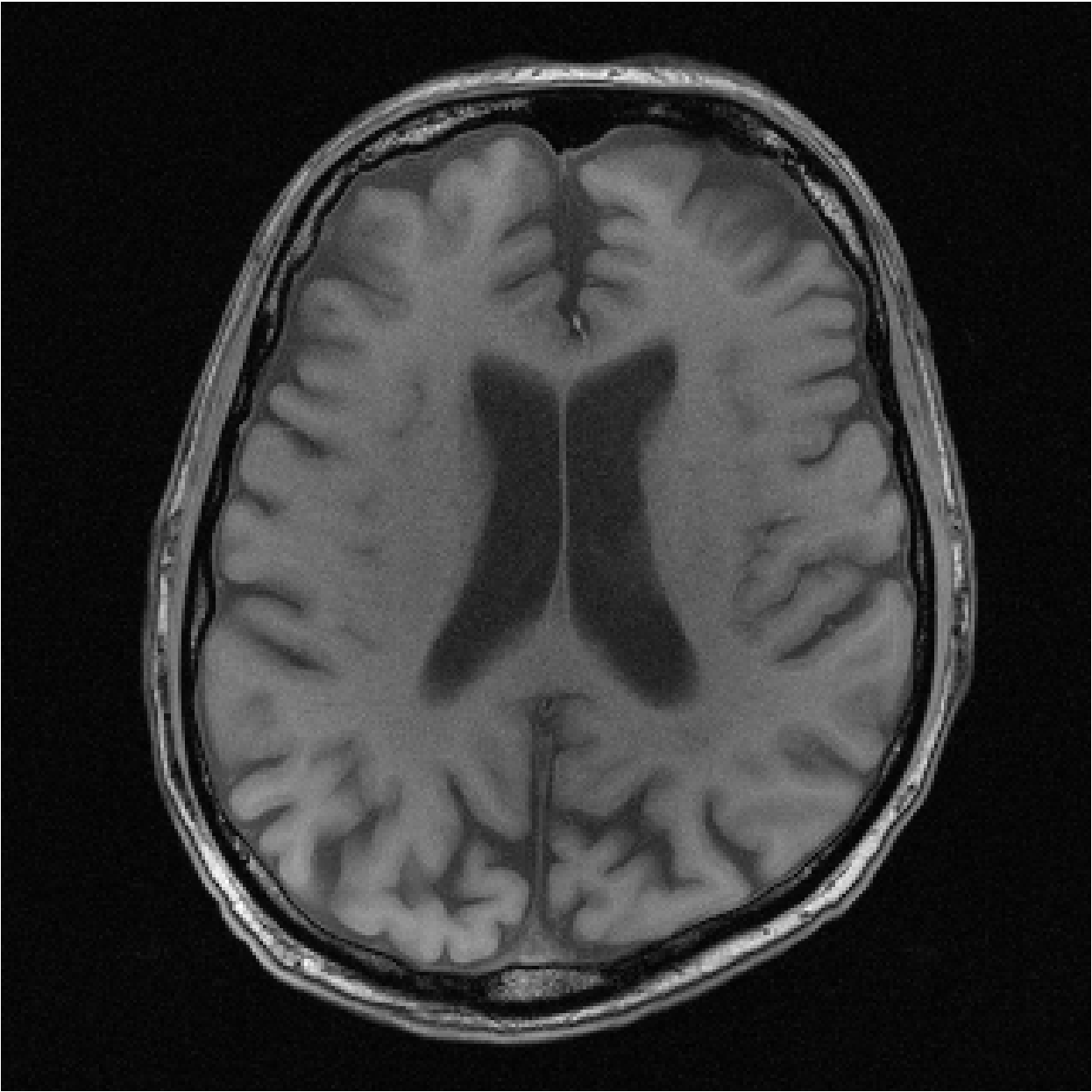}\hfill
\includegraphics[width=0.3333333\columnwidth,trim = 0cm 0cm 0cm 0cm,clip]{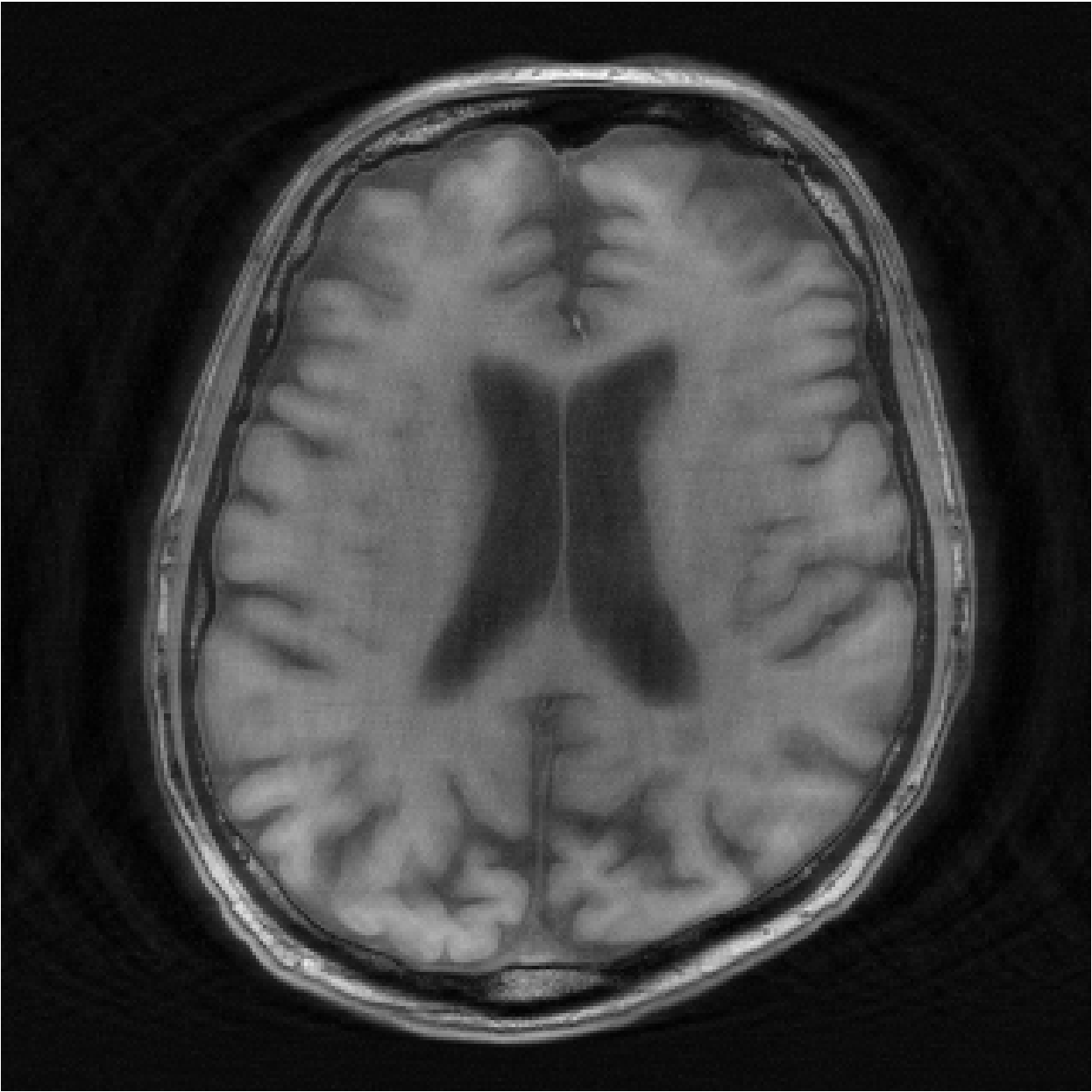}\hfill
\includegraphics[width=0.3333333\columnwidth,trim = 0cm 0cm 0cm 0cm,clip]{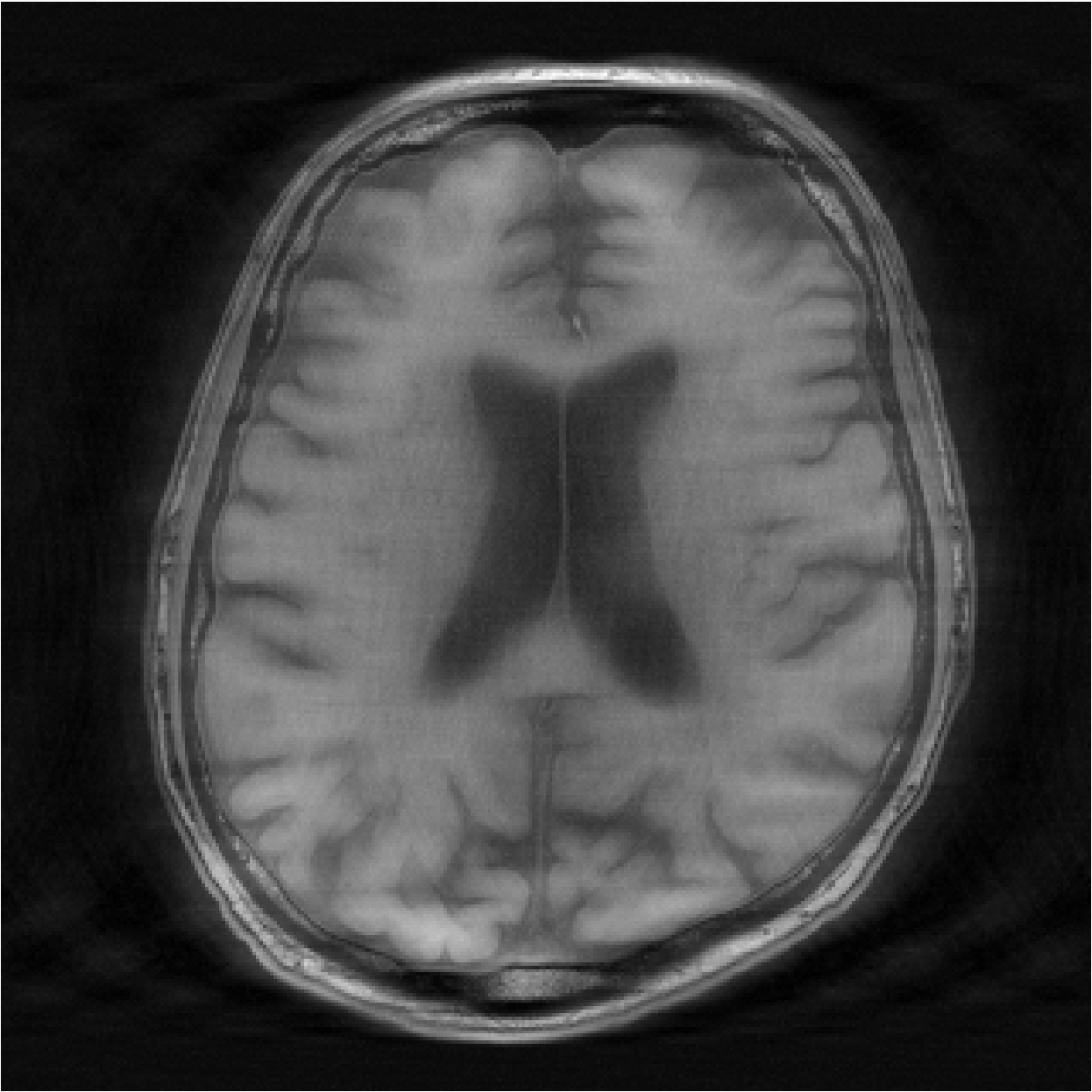}\\
\hspace{2.82cm}
\includegraphics[width=0.3333333\columnwidth,trim = 0cm 0cm 0cm 0cm,clip]{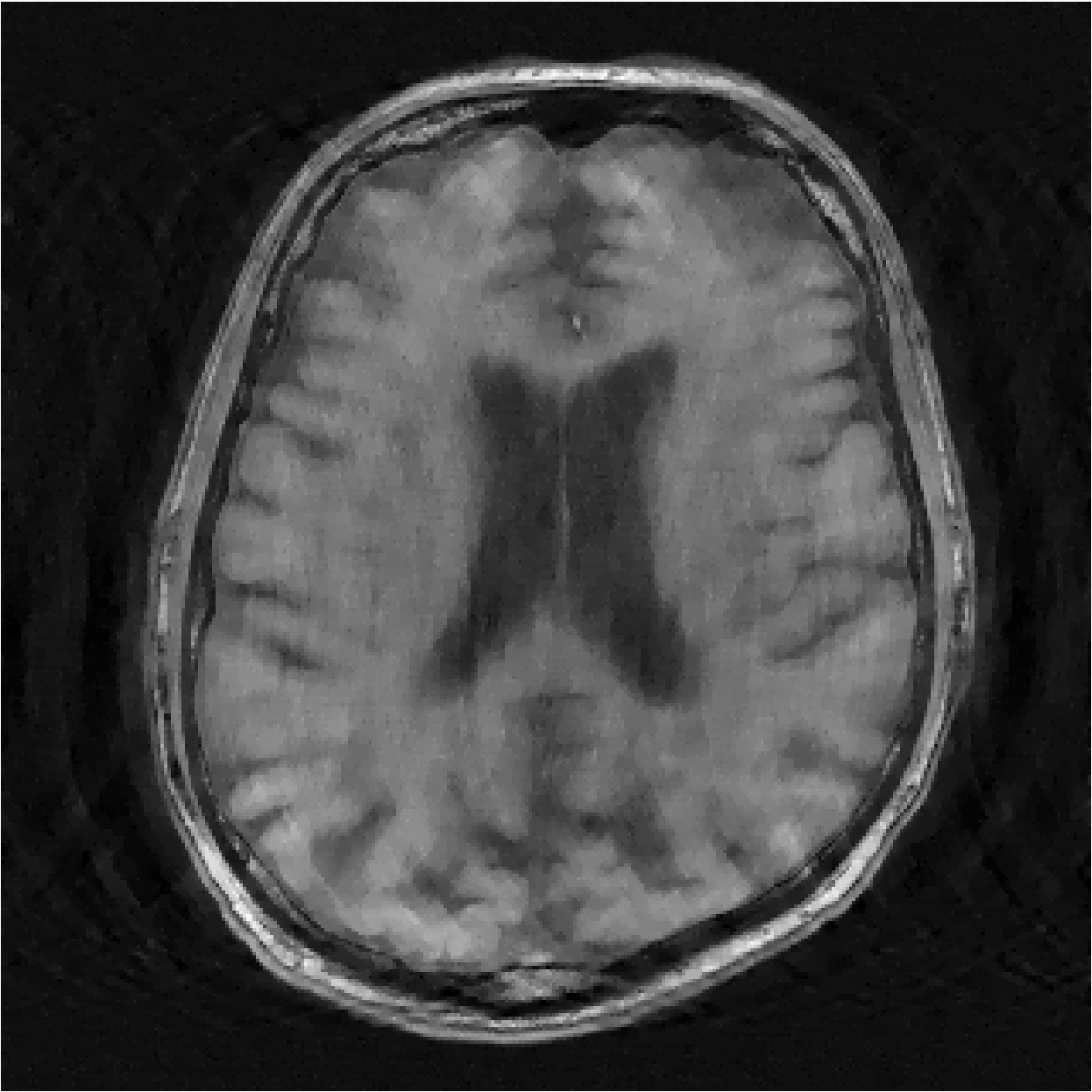}\hfill
\includegraphics[width=0.3333333\columnwidth,trim = 0cm 0cm 0cm 0cm,clip]{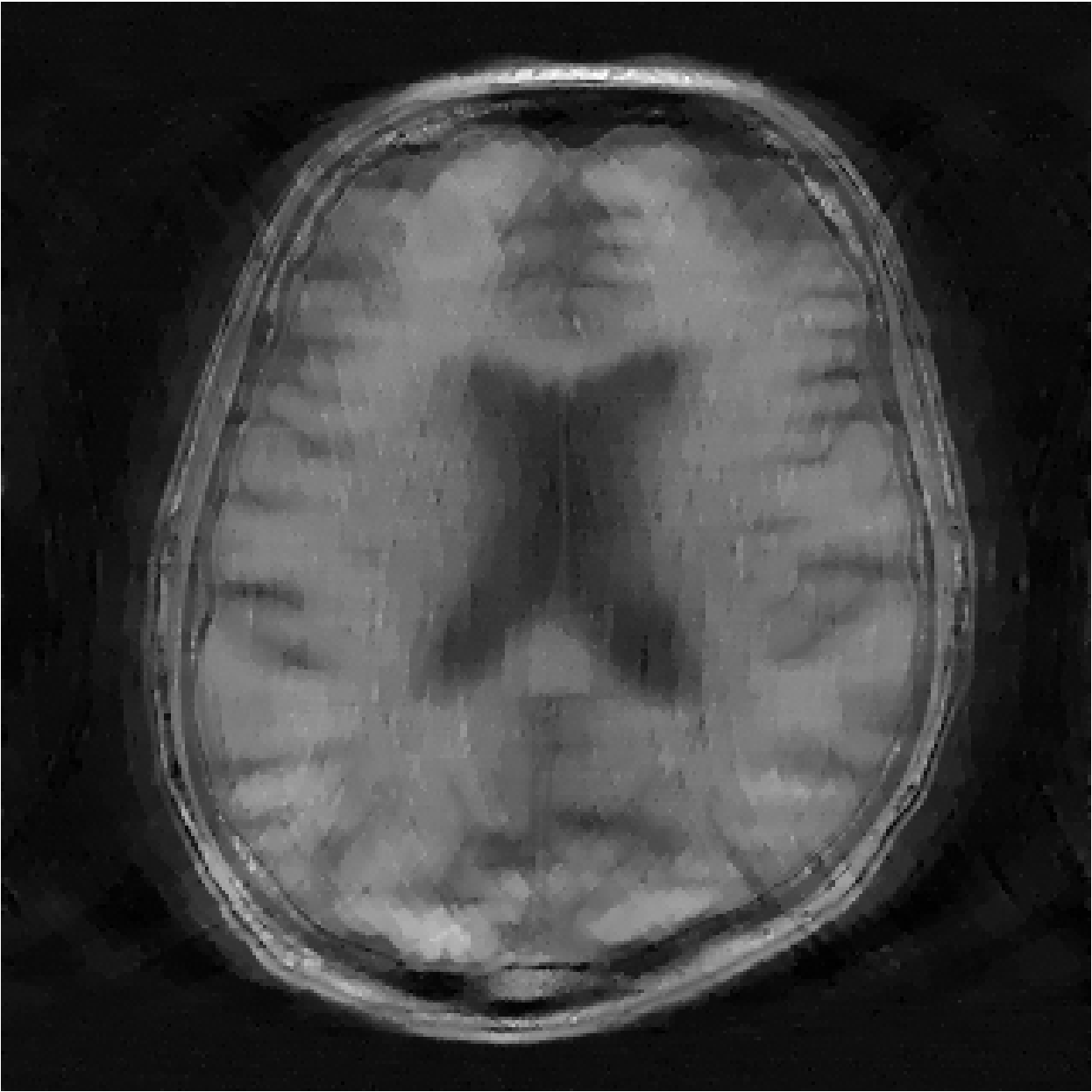}
\caption{Visual comparison between reconstruction of knee and brain images \emph{A} from accelerated MRI measurements. Left column: Fully sampled image. Middle column: Reconstruction from 4-fold under-sampling with our proposed method (top rows) and TV-method (bottom rows).
Right column: Reconstruction from 8-fold under-sampling with our proposed method (top rows) and TV-method (bottom rows).}
\label{figs:A}
\end{figure}

\begin{figure}[h!]
\centering
\includegraphics[width=0.3333333\columnwidth,trim = 0cm 0cm 0cm 0cm,clip]{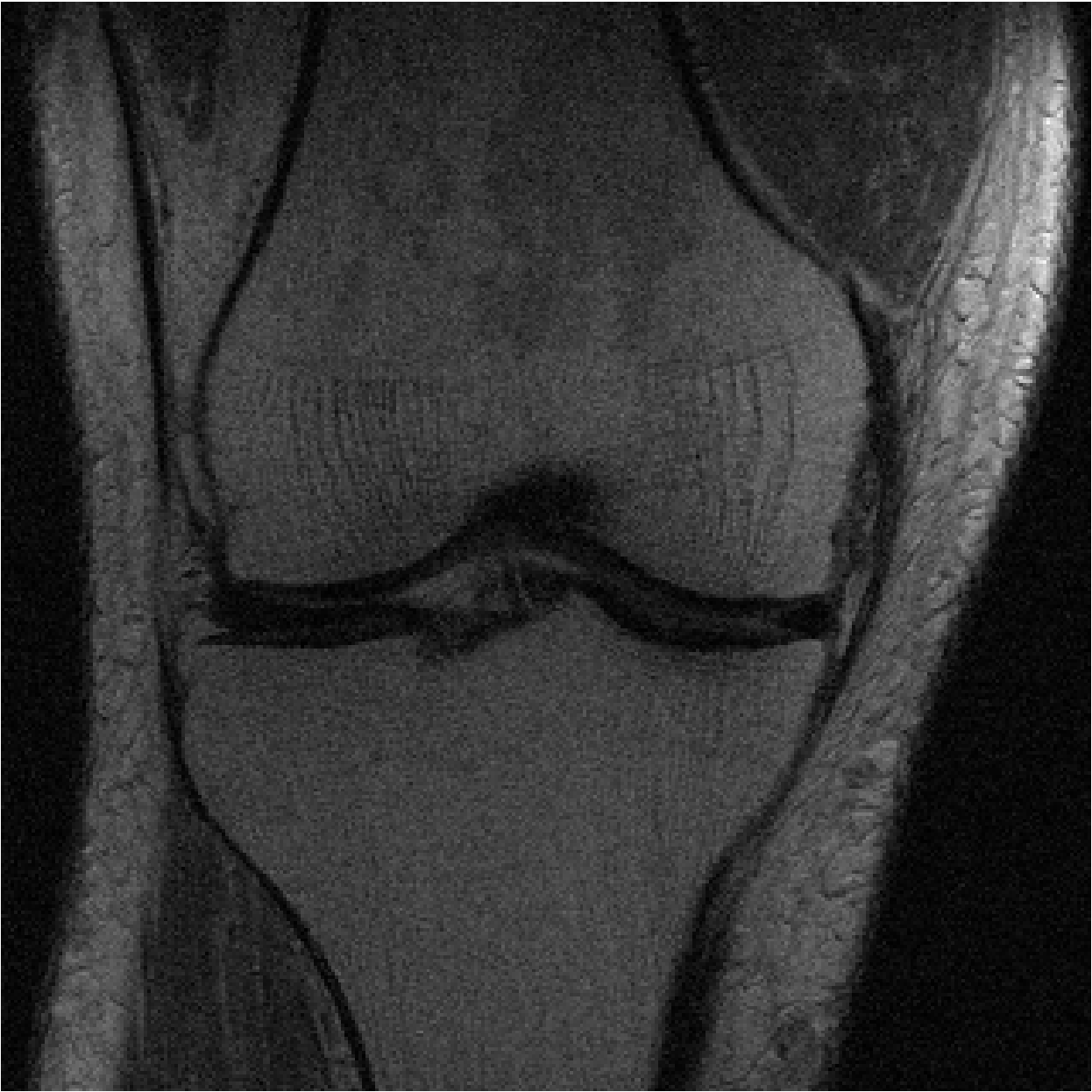}\hfill
\includegraphics[width=0.3333333\columnwidth,trim = 0cm 0cm 0cm 0cm,clip]{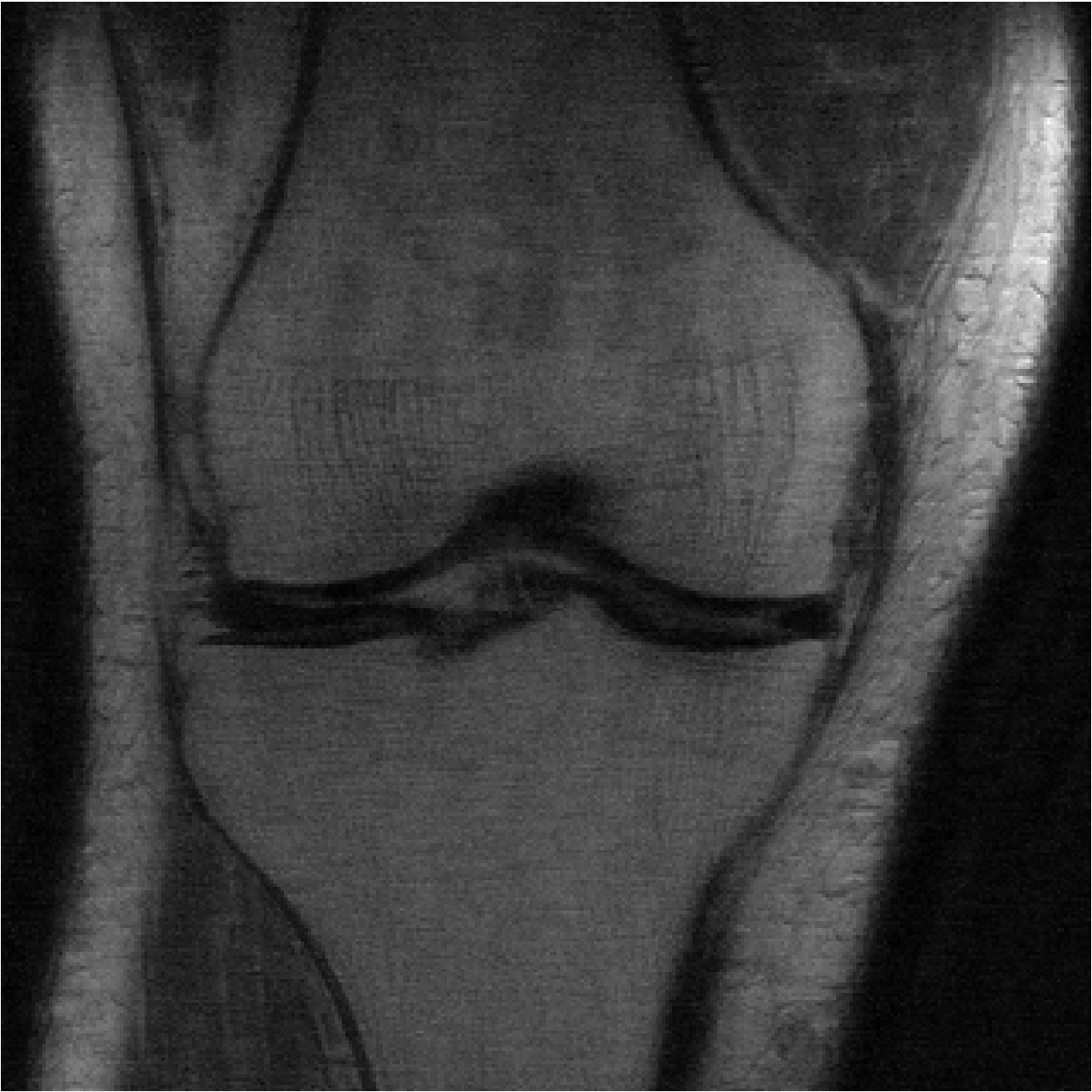}\hfill
\includegraphics[width=0.3333333\columnwidth,trim = 0cm 0cm 0cm 0cm,clip]{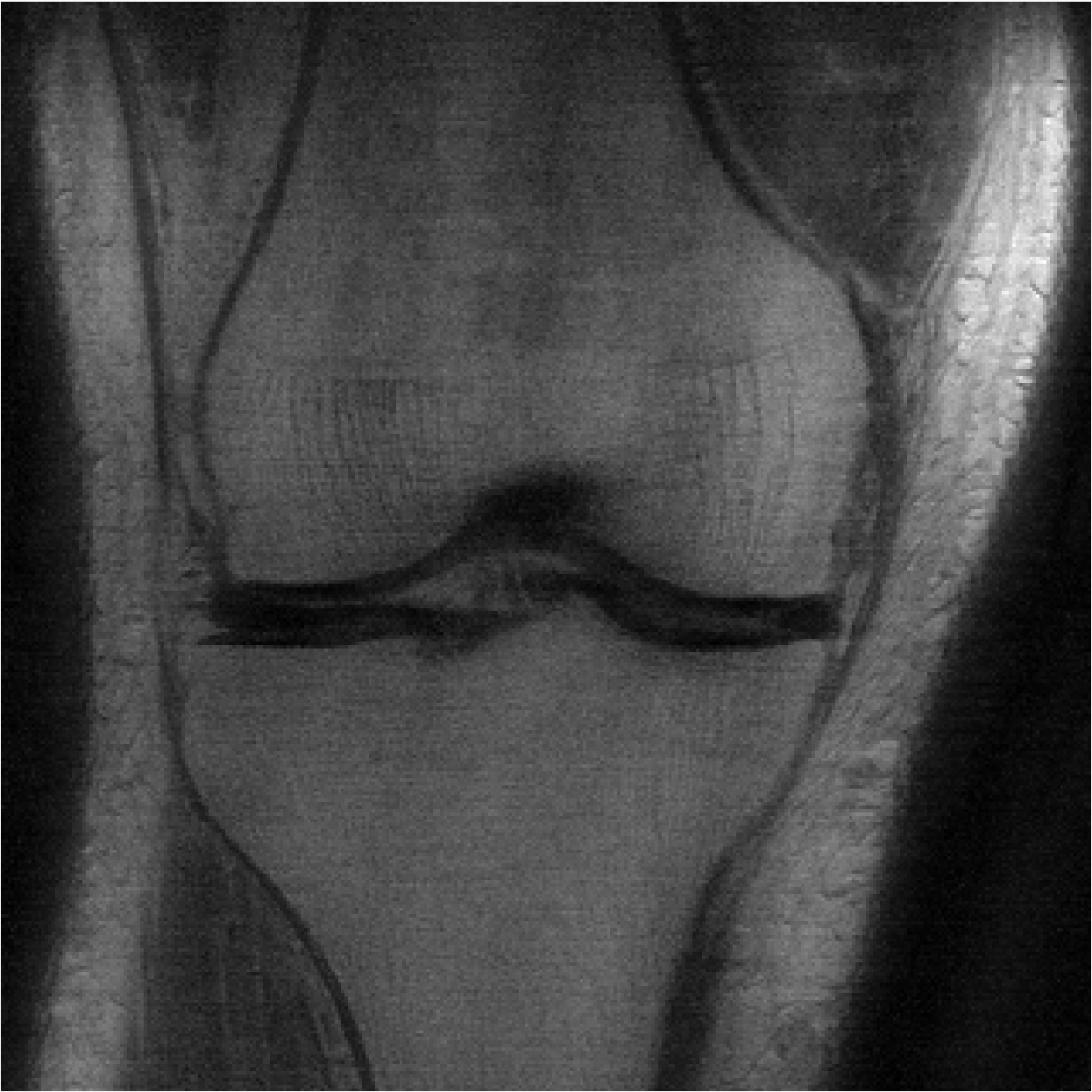}\\
\hspace{2.82cm}
\includegraphics[width=0.3333333\columnwidth,trim = 0cm 0cm 0cm 0cm,clip]{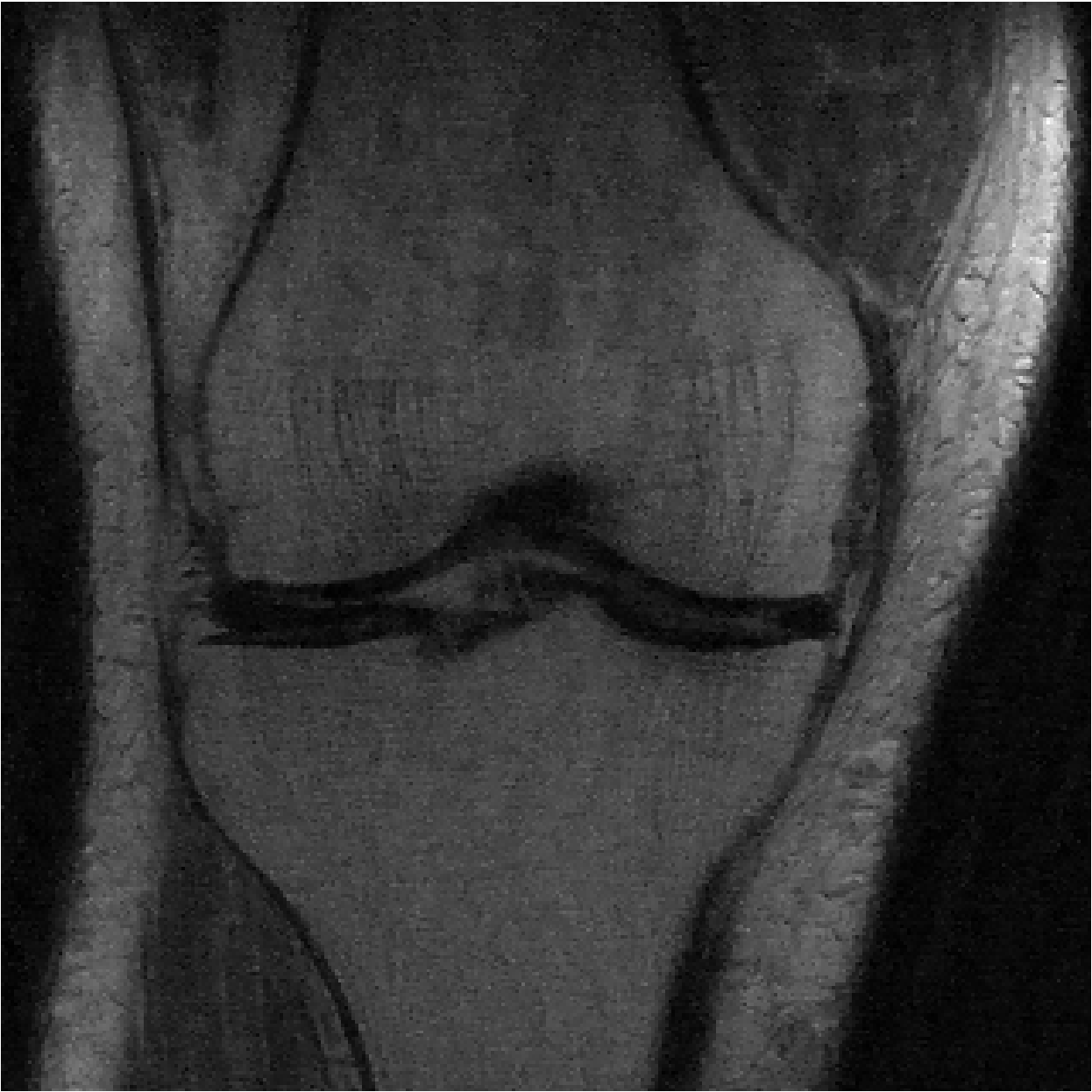}\hfill
\includegraphics[width=0.3333333\columnwidth,trim = 0cm 0cm 0cm 0cm,clip]{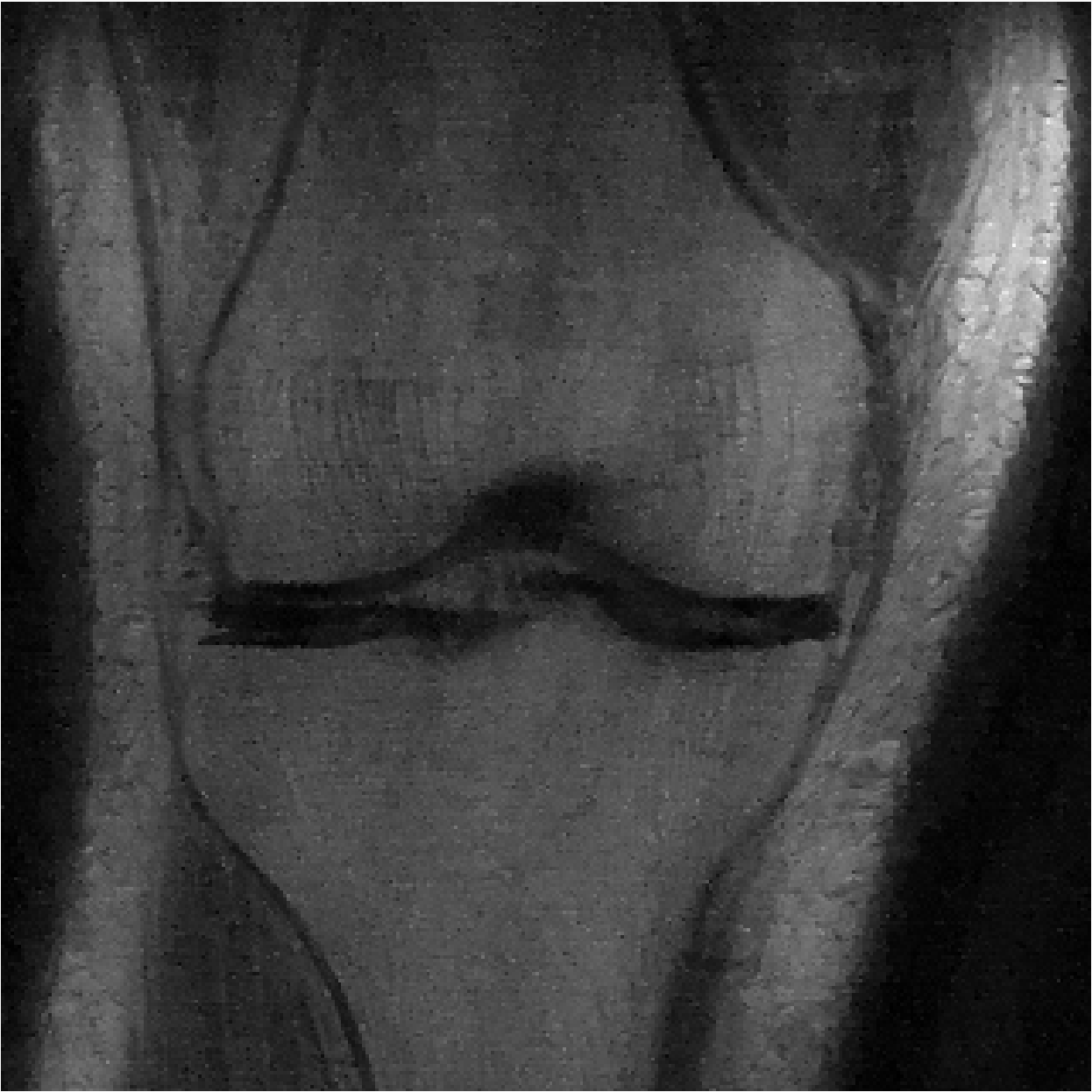}\\
\vspace{0.1cm}
\includegraphics[width=0.3333333\columnwidth,trim = 0cm 0cm 0cm 0cm,clip]{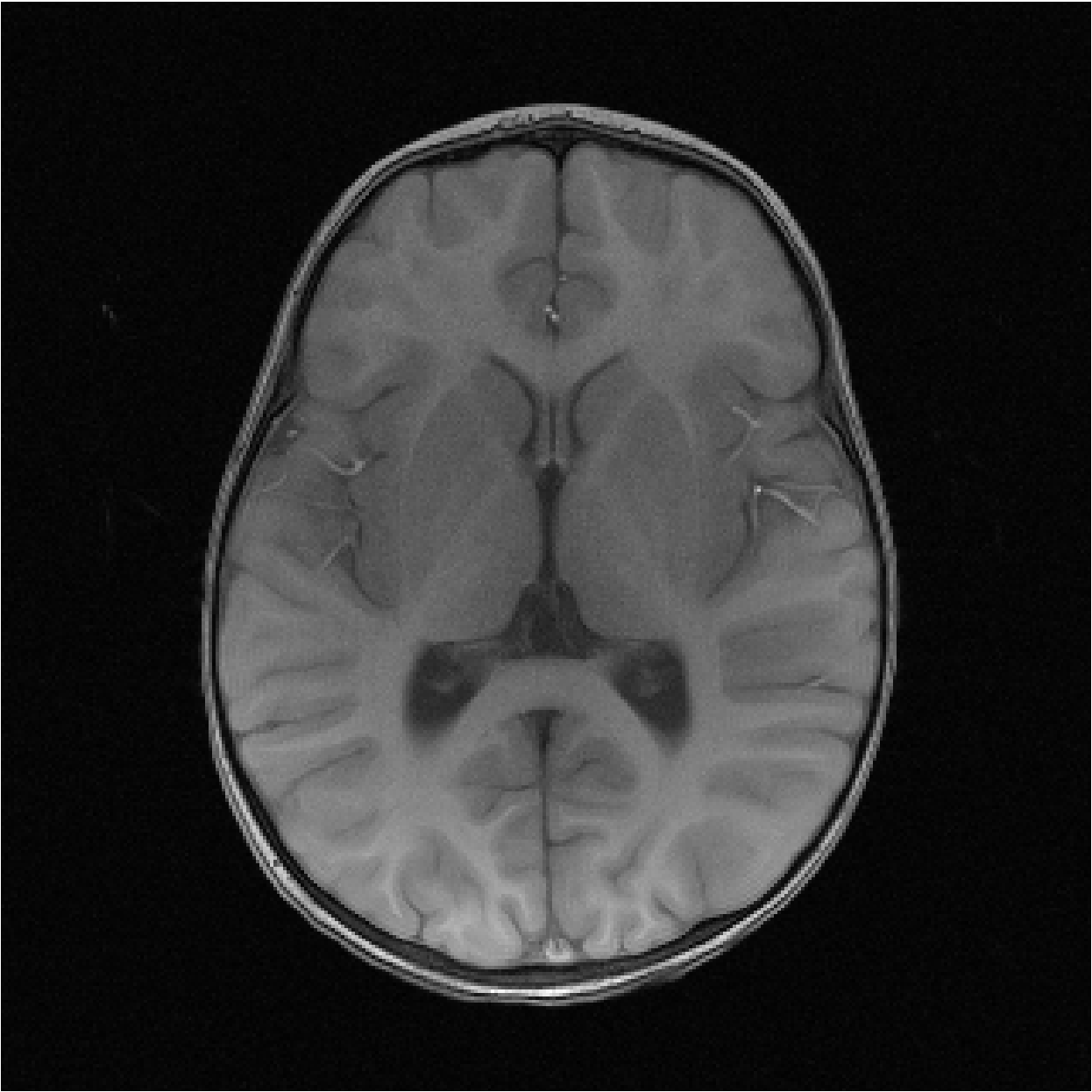}\hfill
\includegraphics[width=0.3333333\columnwidth,trim = 0cm 0cm 0cm 0cm,clip]{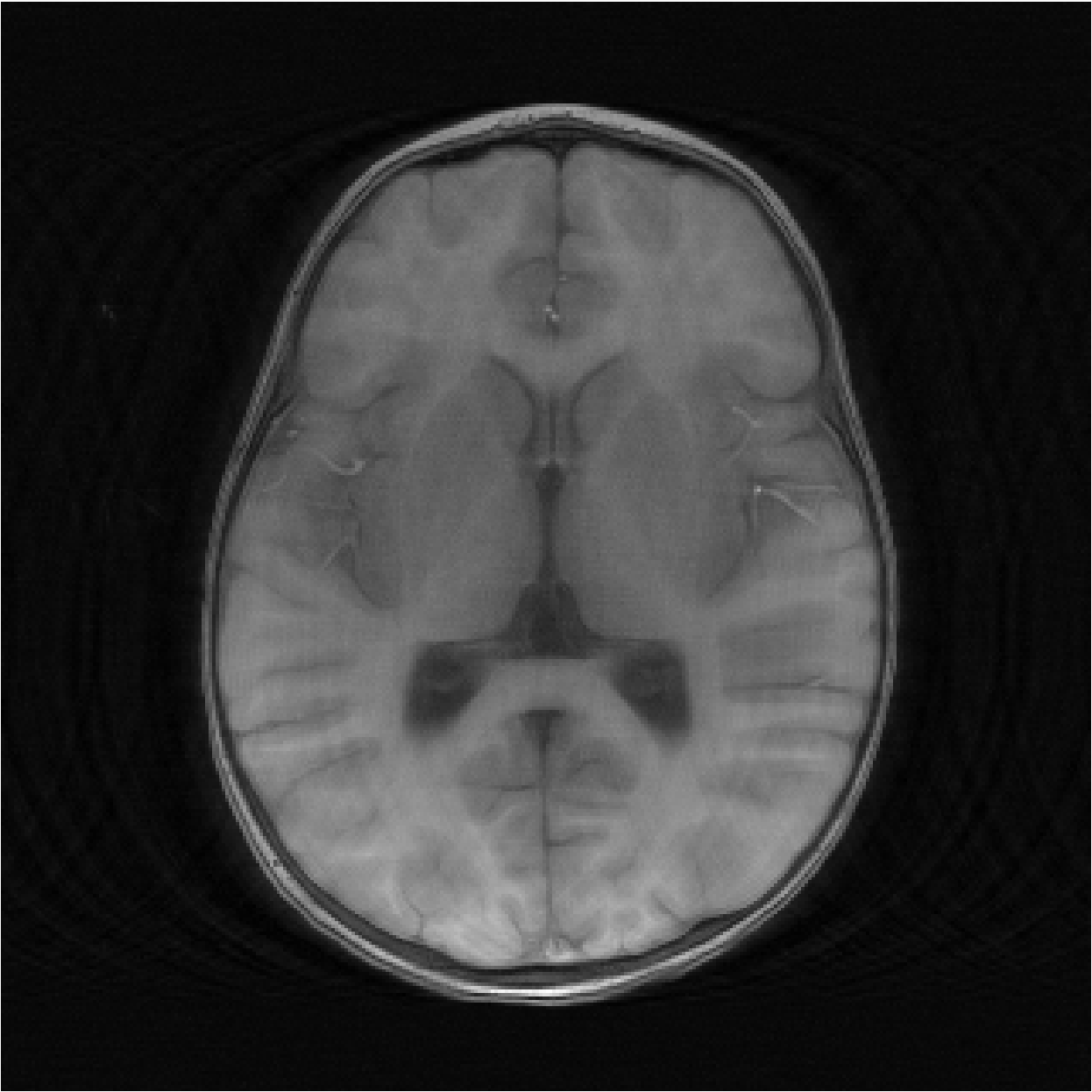}\hfill
\includegraphics[width=0.3333333\columnwidth,trim = 0cm 0cm 0cm 0cm,clip]{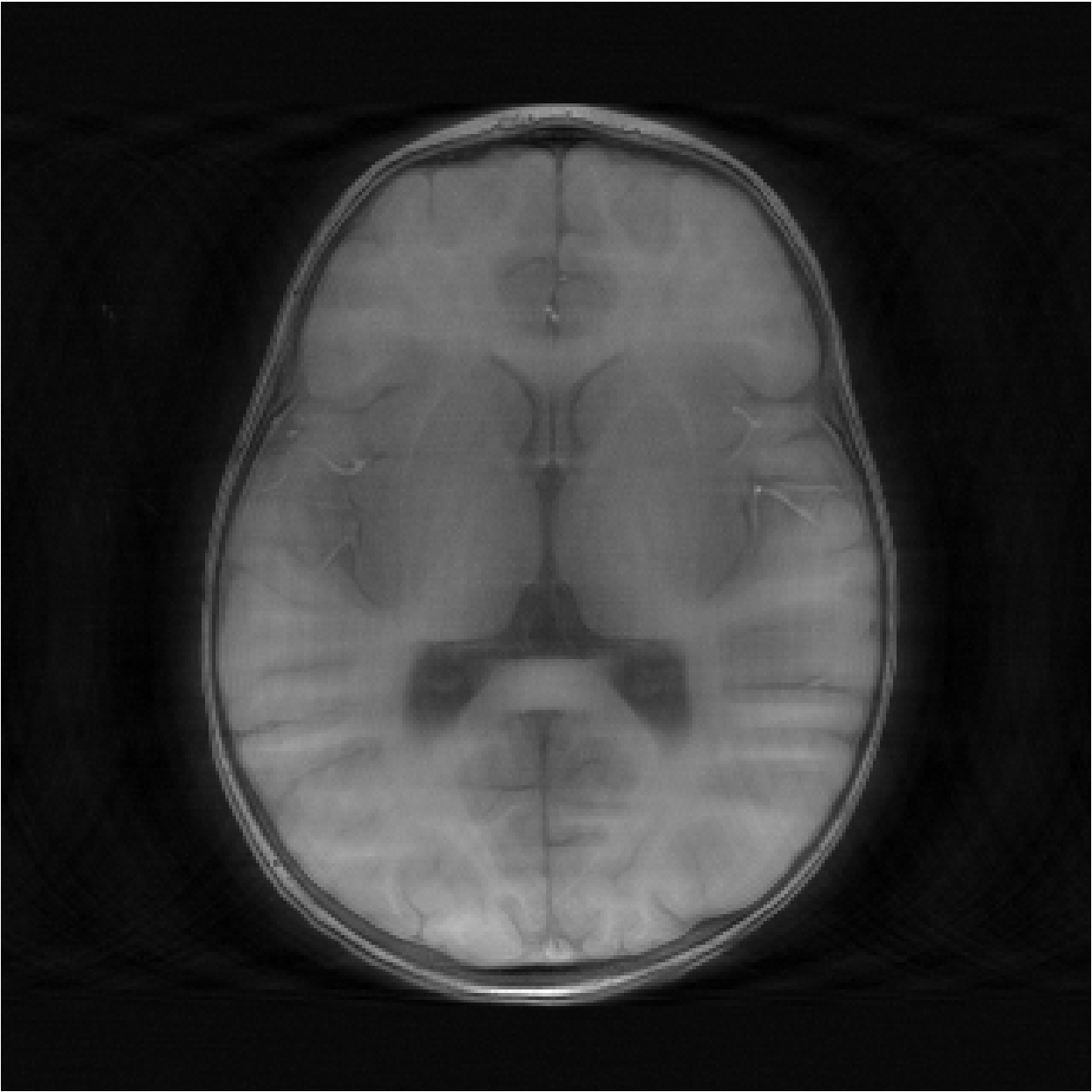}\\
\hspace{2.82cm}
\includegraphics[width=0.3333333\columnwidth,trim = 0cm 0cm 0cm 0cm,clip]{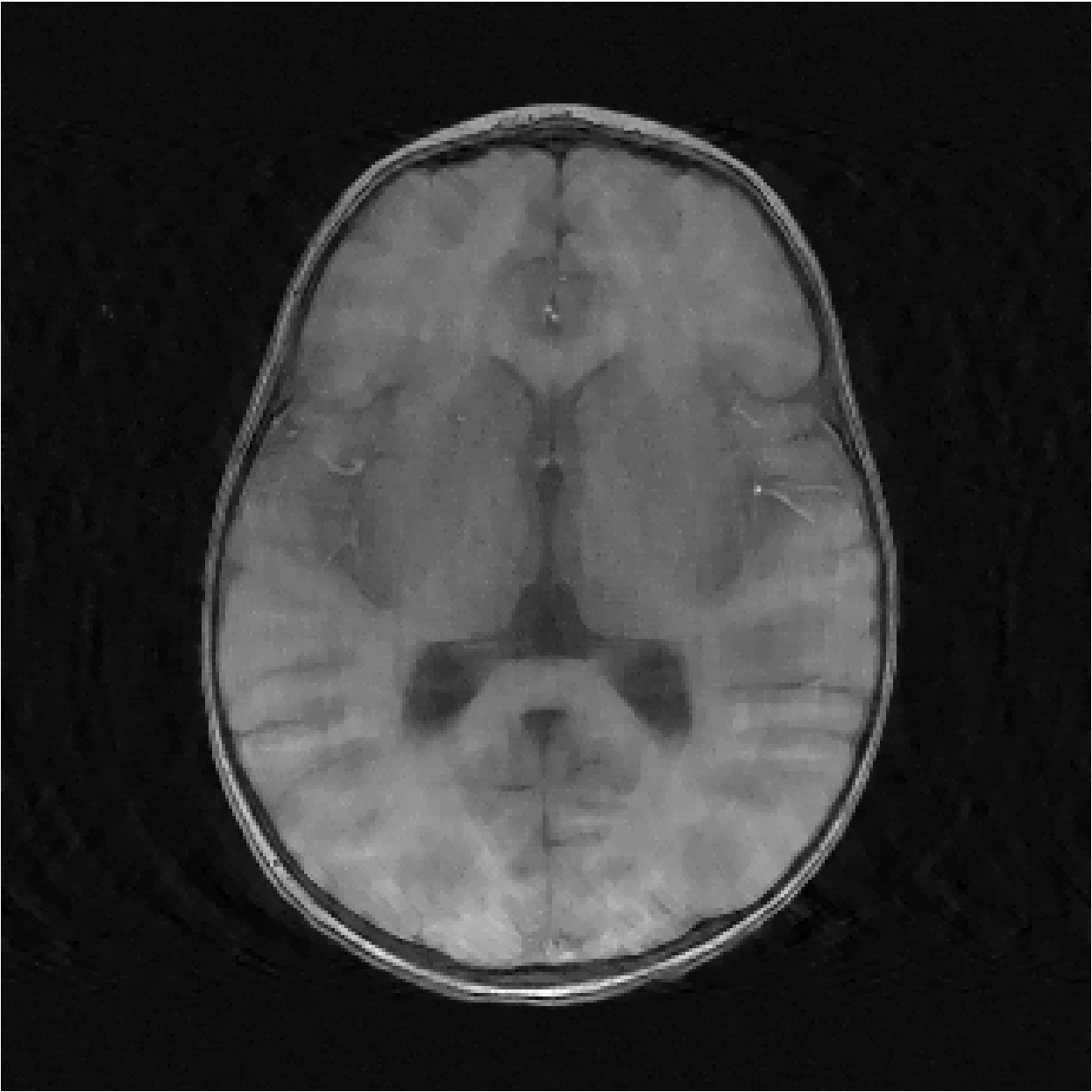}\hfill
\includegraphics[width=0.3333333\columnwidth,trim = 0cm 0cm 0cm 0cm,clip]{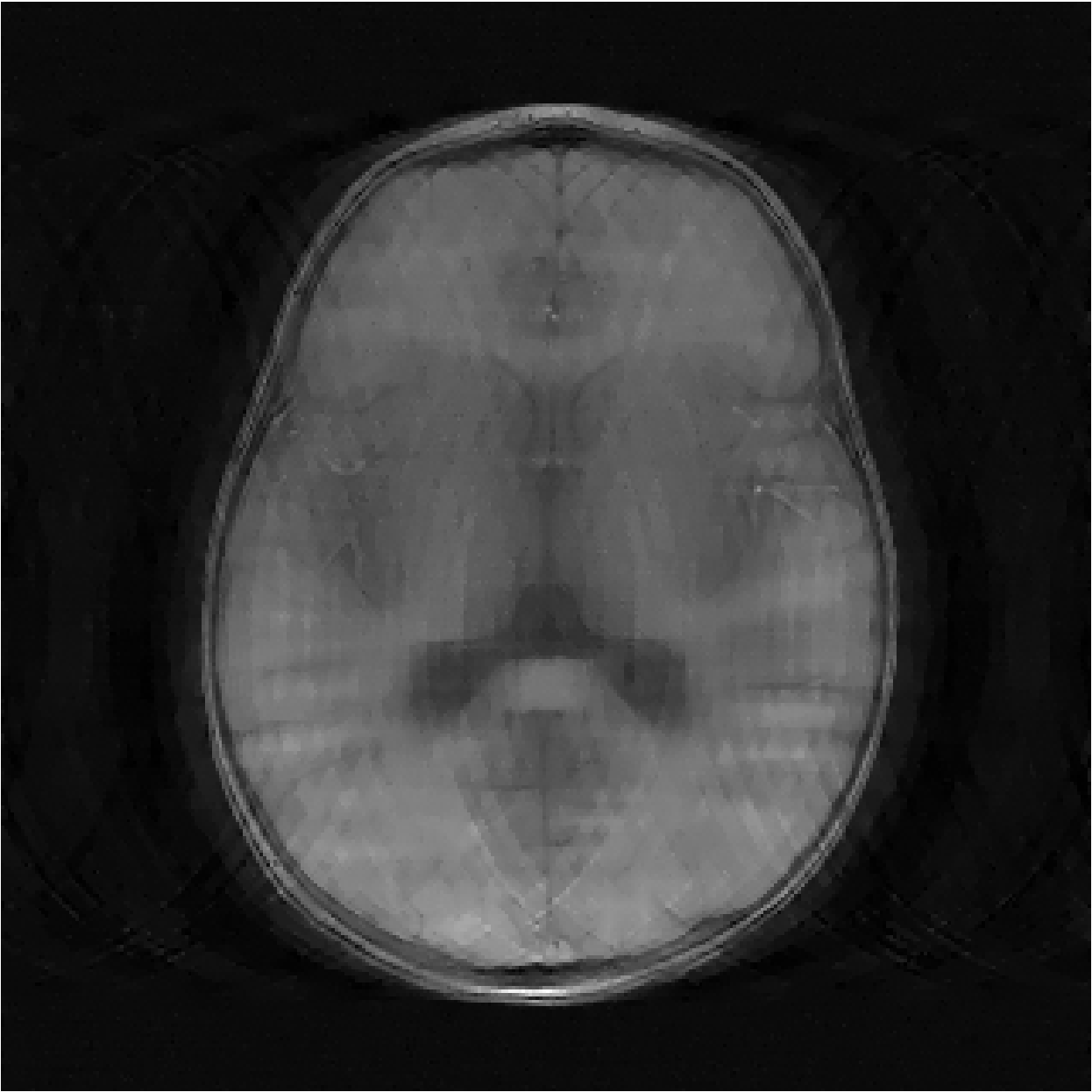}
\caption{Visual comparison between reconstruction of knee and brain images \emph{B} from accelerated MRI measurements. Left column: Fully sampled image. Middle column: Reconstruction from 4-fold under-sampling with our proposed method (top rows) and TV-method (bottom rows).
Right column: Reconstruction from 8-fold under-sampling with our proposed method (top rows) and TV-method (bottom rows).}
\label{figs:B}
\vspace{-0.5cm}
\end{figure}

Figures~\ref{figs:A} and \ref{figs:B} shows a qualitative comparison of the reconstruction results by our algorithm and the total-variation (TV) method from 4-fold and 8-fold under-sampled Fourier measurements for a knee and a brain scan from the fastMRI dataset. 
Parameters were chosen for best performance from a set of values as
$\rho = 0.02$,  
$\nu=0.00016$, 
$\mu=0.00024$, as well as 
$\lambda=5 \times 10^{-5}$ in~\eqref{TValgorithm}, and $\gamma = 0.0035$ in \eqref{SparseApprox}.
MRI data: \texttt{https://fastmri.med.nyu.edu/}.
Images (displayed with 90 degree rotation and cut to quadratic centre region) taken from coronal proton density knee dataset (image size 372×640, $z$-slice 20 of \texttt{file1000031} (knee $A$) and \texttt{file1000071} (knee $B$)), and brain dataset (image size 320×640, $z$-slice 2 of \texttt{file\_brain\_AXT1\_201\_6002786} (brain $A$) and 
\texttt{file\_brain\_AXT1\_201\_6002740} (brain $B$)).
As shown in Figures~\ref{figs:A} and \ref{figs:B}, the lowpass wavelet band and the center regions of the acceleration masks are not perfectly matched, causing artifacts in the reconstruction due to cross-band interference between the lowpass and highpass wavelet bands. Table \ref{tabl_perf} summarizes a comparison of PSNR and SSIM values for the knee and brain images depicted in Fig.~\ref{figs:A} and Fig.~\ref{figs:B}.  Our average PSNR gains for 4-fold under-sampled data are $0.5$ dB on the knee and $0.9$ dB on the brain images, while our average gains for the 8-fold under-sampled images are about $2$ dB. Our SSIM values are superior to that of the TV-method in all experiments. 

\begin{table}[ht]
\caption{Performance comparison for images in Fig.~\ref{figs:A}, and Fig.~\ref{figs:B} in appendix, for 4-fold  and 8-fold  acceleration.}
\centering
\ra{1.2}
\begin{tabular}[h]{@{}lcccccccc@{}}\toprule 
&   \multicolumn{4}{c}{Proposed}  &  \multicolumn{4}{c}{TV method}\\
\cmidrule(lr){2-5}\cmidrule(lr){6-9}
 &  \multicolumn{2}{c}{PSNR} & \multicolumn{2}{c}{SSIM} & \multicolumn{2}{c}{PSNR} & \multicolumn{2}{c}{SSIM}\\
 \cmidrule(lr){2-3}\cmidrule(lr){4-5} \cmidrule(lr){6-7}\cmidrule(lr){8-9}
 & 4x& 8x & 4x& 8x & 4x& 8x & 4x& 8x\\
 \midrule
 Knee \emph{A}  & 28.5&26.4 & 0.657& 0.570 & 28.2 & 24.5& 0.615& 0.538\\
  Knee \emph{B}  & 28.4&26.1 & 0.635& 0.535 & 27.8 & 24.1& 0.625& 0.541\\
   Brain \emph{A}  & 27.7&24.1 & 0.642& 0.524 & 26.7 & 22.3& 0.637& 0.493\\
    Brain \emph{B}  & 27.7&24.6 & 0.668& 0.494 & 26.8 & 22.7& 0.644& 0.474\\
\bottomrule
\end{tabular}
\label{tabl_perf}
\end{table}

\section{Conclusions} \label{conclusion}
This article describes a novel approach to CS involving the construction a sensing matrix for a prescribed sparsifying dictionary, such that their composition has the RIP. We use a matrix factorization of the dictionary into an invertible, a random and an orthonormal factor. The random matrix satisfies a CS concentration inequality and thus possess the RIP with high probability, which then transfers to the composition of sensing matrix and dictionary. 
We further apply the factorization approach to accelerated MR imaging - deriving an embedding for under-sampled $k$-space data that facilitates RIP recovery guarantees. In this application one of the factors in our proposed decomposition is adapted to the physically prescribed sensing mechanism, which otherwise does not allow for direct RIP based recovery guarantees in the $\ell_1$-synthesis approach to CS. 
A future direction is to leverage our approach with the technique of \emph{unrolling} \cite{gu2022revisiting} regularization-based CS methods into a novel architecture neural net, while matching the performance of large baseline deep neural nets for accelerated MRI.

\bibliographystyle{ieeetr}
\bibliography{MRI_CSref}

\newpage

\appendix

To solve \eqref{opti_G} numerically via the Lagrangian approach we consider the augmented Lagrangian
\begin{align*}\label{Largr_G}
L_1(G,\widetilde{G}) &:= 
\tfrac{1}{2}\|\mathcal{R}(\Real(F_1)G -I)\|_F^2\\
&\quad+ \text{Tr}\left(\lambda_1^{\top} (\widetilde{G} G -I)\right)
+ \text{Tr}\left(\lambda_2^{\top} ( G \widetilde{G} -I)\right) \\
&\quad+\tfrac{\rho}{2}\left(\|\widetilde{G} G -I \|_F^2 + \|G \widetilde{G} -I \|_F^2\right),
\end{align*}
with fixed $\rho>0$. We initialize $G_R$ as in \eqref{rangeConstructions}, $\widetilde{G}$ as $G_R^{-1}$, and $\lambda_1=\lambda_2$ as zero matrices, and iterate the updates
\begin{enumerate}
\item[(i)] $G_R \leftarrow \argmin_{G}  L_1(G,\widetilde{G})$
\item[(ii)] $\widetilde{G} \leftarrow \argmin_{\widetilde{G}}  L_1(G_R,\widetilde{G}) $ 
\item[(iii)] 
$\lambda_1 \leftarrow \lambda_1 + \rho(\widetilde{G} {G_R} -I)$ and 
$\lambda_2 \leftarrow \lambda_2 + \rho( {G_R} \widetilde{G} -I)$ 
\end{enumerate}
\noindent 
until convergence of  $\lambda_1$ and $\lambda_2$.
Steps (i) and (ii) can be approximated by solving the Sylvester equations $\nabla_G L_1 = 0 $ and $\nabla_{\widetilde{G}} L_1 = 0$ numerically (e.g., via matlab function sylvester), where 
\begin{align*}
\nabla_G L_1
&= 
 \Real(F_1)^{\top} \mathcal{R}(\Real(F_1)G -I)
+ \widetilde{G}^{\top} \lambda_1
+  \lambda_2 \widetilde{G}^{\top}  \\
&\quad+\rho ( \widetilde{G}^{\top} (\widetilde{G} G-I ) 
+ (G  \widetilde{G}  -I ) \widetilde{G}^{\top}  )\\
&= ( \Real(F_1)^{\top} \mathcal{R} \Real(F_1)+  \rho \widetilde{G}^{\top} \widetilde{G}) G 
+ G  ( \rho\widetilde{G} \widetilde{G}^{\top} ) \\
&\quad
+ ( \widetilde{G}^{\top} \lambda_1
+  \lambda_2 \widetilde{G}^{\top} 
-\Real(F_1)^{\top} \mathcal{R} - 2\rho \widetilde{G}^{\top}),
\end{align*}
and 
\begin{align*}
\nabla_{\widetilde{G} }  L_1
&=
G^{\top} \lambda_2 
+  \lambda_1 G^{\top} 
+\rho ( (\widetilde{G} G  -I ) G^{\top}
+G^{\top} (G  \widetilde{G}  -I )  )\\
&= \rho (G^{\top} G \widetilde{G} 
+ \widetilde{G}  G G^{\top} ) 
+ G^{\top} \lambda_2
+  \lambda_1 G^{\top}   - 2\rho G^{\top}.
\end{align*}

Problem \eqref{opti_H} is solved analogously, now considering
\begin{align*}
L_2(H,\widetilde{H}):= &  \tfrac{1}{2}\| D_1 - G_RAH \|_F^2
 + \tfrac{\nu}{2}\| H- \widetilde{H}^{\top}  \|_F^2 \\
&+ \text{Tr}\left(\lambda_3^{\top} (\widetilde{H} H -I)\right) + \text{Tr}\left(\lambda_4^{\top} ( H \widetilde{H} -I)\right) \\
&+\tfrac{\mu}{2}\left(\|\widetilde{H} H -I \|_F^2 + \|H \widetilde{H} -I \|_F^2\right),
\end{align*}
initializing $H_R$ as in \eqref{rangeConstructions}, $\widetilde{H}$ as $H_R^\top$ and $\lambda_3=\lambda_4$ as zero matrices. The iteration steps are now 
\begin{enumerate}
\item[(i')] $H_R \leftarrow \argmin_H  L_2(H,\widetilde{H})$
\item[(ii')] $\widetilde{H}\leftarrow \argmin_{\widetilde{H}}  L_2(H_R,\widetilde{H})$
\item[(iii')]
$\lambda_3 \leftarrow \lambda_3 + \mu(\widetilde{H} H_R -I)$ and 
$\lambda_4 \leftarrow \lambda_4 + \mu(H_R\widetilde{H} -I)$, 
\end{enumerate}
and the derivatives concerning (i') and (ii') are
%
\begin{align*}
\nabla_H  L_2  
&=
A^{\top} G^{\top} (D - GAH) 
+ \nu  \widetilde{H}( H- \widetilde{H}^{\top} ) \\
&\quad
+ \widetilde{H}^{\top} \lambda_3
+ \lambda_4 \widetilde{H}^{\top} \\
&\quad+ \mu (\widetilde{H}^{\top} ( \widetilde{H} H  -I ))
+ \mu ( ( H \widetilde{H} -I )\widetilde{H}^{\top}) \\
& =  (A^{\top} G^{\top} GA + \nu  \widetilde{H} + \mu \widetilde{H}^{\top} \widetilde{H})H 
+ H(\mu \widetilde{H} \widetilde{H}^{\top}) \\
&\quad +  A^{\top} G^{\top} D +\widetilde{H}^{\top} \lambda_3 + \lambda_4 \widetilde{H}^{\top} 
-2\mu  \widetilde{H}^{\top} - \nu \widetilde{H} \widetilde{H}^{\top},
\end{align*}
and
\begin{align*}
\nabla_{\widetilde{H} }  L_2 
&= 
\nu  H( \widetilde{H}- H^{\top} )
+ \lambda_3 {H}^{\top} 
+ {H}^{\top}\lambda_4  \\
&\quad+ \mu ( ( \widetilde{H} H  -I ) {H}^{\top})
+ \mu ( {H}^{\top}( H \widetilde{H} -I )) \\
&= ( \nu  {H} + \mu {H}^{\top} {H})\widetilde{H}
+ \widetilde{H}(\mu {H} {H}^{\top}) \\
&\quad+ (\lambda_3 {H}^{\top} 
+ {H}^{\top}\lambda_4
-2\mu  {H}^{\top} - \nu {H} {H}^{\top}).
\end{align*}

\end{document}